\documentclass{JHEP3}

\usepackage{epsfig,multicol,bbm}
\usepackage{graphicx}
\usepackage{dcolumn}
\usepackage{amsmath}
\usepackage{enumerate}

\newcommand{\Cc}{{\cal C}}

\newcommand{\CC}{{\bf C}}

\title{Cosmological parameters degeneracies and non-Gaussian halo bias}

\author{Carmelita Carbone \\ Dipartimento di Astronomia, Universit\a`a di Bologna\\
  Via Ranzani 1, I-40127 Bologna, Italy \\ E-mail: \email{carmelita.carbone@unibo.it}}

\author{Olga Mena\\ Depto.\ de F\'{\i}sica Te\'orica, IFIC, Universidad de
Valencia-CSIC \\ Edificio de Institutos de Paterna, Apt. 22085, 46071 Valencia,
Spain\\ E-mail: \email{omena@ific.uv.es}}

\author{Licia Verde\\
ICREA \& Instituto de Ciencias del Cosmos, Universitat de Barcelona \\
Marti i Franques 1, 08028, Barcelona, Spain \\
E-mail: \email{liciaverde@icc.ub.edu}}

\preprint{\tt arXiv:1003.0456}

\abstract{
{\small We study the  impact of the cosmological parameters uncertainties
on the measurements of primordial  non-Gaussianity  through the  
large-scale non-Gaussian halo bias effect.
While this is not expected to be an issue for the standard $\Lambda$CDM  
model, it may not be the case for more general models that modify the  
large-scale shape of the power spectrum.  We consider the so-called  
local non-Gaussianity model, parametrized by the $f_{\rm NL}$
  non-Gaussianity parameter which is zero for a Gaussian case, and
make forecasts on $f_{\rm NL}$ from planned surveys, alone  
and combined with a Planck CMB prior.
In particular, we consider EUCLID- and LSST-like surveys and forecast  
the correlations among $f_{\rm NL}$ and
the running of the spectral index $\alpha_s$, the dark energy
equation of state $w$, the effective sound speed of dark energy  
perturbations $c^2_s$, the total mass of
massive neutrinos $M_\nu=\sum m_\nu$, and the number of
extra relativistic degrees of freedom $N_\nu^{rel}$.
Neglecting CMB information on $f_{\rm NL}$
and scales $k > 0.03 h$/Mpc, we find that, if $N_\nu^{\rm rel}$ is
assumed to be known, the uncertainty on cosmological parameters  
increases the error on $f_{\rm NL}$ by 10 to 30\% depending on the survey.  Thus  
the $f_{\rm NL}$ constraint is remarkable robust to cosmological  
model uncertainties.
On the other hand, if $N_\nu^{\rm rel}$ is simultaneously constrained  
from the data, the $f_{\rm NL}$  error increases by $\sim 80\%$.
Finally, future surveys which
provide a large sample of galaxies or galaxy clusters over a volume
comparable to the Hubble volume can measure primordial
non-Gaussianity of the local form with a marginalized 1--$\sigma$
error of the order $\Delta f_{\rm NL} \sim 2-5$, after combination  
with CMB priors for the remaining cosmological parameters.
These results are competitive with CMB bispectrum constraints
achievable with an ideal CMB experiment.}}

\keywords{cosmology: theory, large-scale structure of universe -- 
galaxies: clusters, general -- galaxies: halos}

\begin{document}
\section{Introduction}
\label{Introduction}
Tests of deviations from Gaussian initial conditions offer an
important window into the very early Universe and a powerful test for
the mechanism which generated primordial perturbations.
While standard single-field slow-roll models of inflation lead to
small departures from Gaussianity, non-standard scenarios allow for a
larger non-Gaussianity (NG) level (e.g. \cite{BKMR04, bartolofnl05, Chenreview}, and
refs. therein).

In particular, large NG can be produced if any of the
conditions below is violated:
{\it a)} single field, {\it b)} canonical kinetic energy {\it c)}
slow roll and {\it d)} adiabatic (Bunch-Davies) initial vacuum state.
The type of NG arising in standard inflation reads
\cite{SalopekBond90, Ganguietal94,VWHK00, KS01}
\begin{equation}
\Phi=\phi+f_{\rm NL}\left(\phi^2-\langle \phi^2 \rangle\right)\,,
\label{eq:fnl}
\end{equation}
where $\Phi$ denotes Bardeen's gauge-invariant potential, which,
on sub-Hubble scales reduces to the usual Newtonian peculiar
gravitational potential up to a minus sign, and $\phi$ denotes a
Gaussian random field.
The NG parameter $f_{\rm NL}$ is
often considered to be a constant, yielding NG of the {\it local} type 
with a bispectrum  which is maximized for squeezed configurations \cite{Creminellishapes}.
NG of the local type is generated in
standard inflationary scenarios (where $f_{\rm NL}$ is expected to be
of the same order of the slow-roll parameters) as well as in multi-field
inflationary scenarios\footnote{Note that Eq.~(\ref{eq:fnl}) is not
general, i.e. there is a plethora of possible deviations from
Gaussianity arising in the different inflationary scenarios proposed in the
literature.}.

The standard observables to constrain NG are the Cosmic Microwave
Background (CMB) and the Large-Scale Structure (LSS) of the Universe.
Traditionally, the most popular method to detect primordial
NG has been to measure the bispectrum or the three-point correlation
function of the CMB \cite{VWHK00, komatsuetal05, yadav}, while the
LSS bispectrum has been shown to be sensitive to primordial NG only at
high redshift \cite{VWHK00, Scocc, sefusattikomatsu, Coor06, PPM07}.

Other  powerful techniques to measure NG are based on weak lensing
tomography \cite{fedeli09}, Integrated Sachs-Wolfe effect (ISW)
\cite{CVM08,AfshordiTolley08}, abundance \cite{MVJ00, VJKM01,
Loverdeetal07, RB00, RGS00} and clustering \cite{GW86, MLB86}
of rare events such as density peaks, since they trace the tail
of the underlying distribution.

Refs. \cite{DDHS07} and \cite{MV08} (hereafter MV08) showed that
primordial NG
affects the clustering of dark matter halos inducing a scale-dependent
large-scale bias. This effect, which goes under the name of
non-Gaussian
halo bias, is particularly promising, yielding already stringent
constraints from existing data \cite{Slosaretal08, AfshordiTolley08}.
Forthcoming constraints on NG exploiting the non-Gaussian halo bias
are expected to be similar to those achievable from an ideal CMB survey
\cite{CVM08}. These predictions have been confirmed by N-body
simulations \cite{DDHS07,Grossietal09,Desjacques,Pillepich2010}.

Forecasts for $f_{\rm NL}$ constraints from the halo-bias have been
carried out so far assuming perfect knowledge 
of all other  cosmological parameters. While for a $\Lambda$CDM model
this is expected to be a reasonable assumption, 
for  more general models one may expect  $f_{\rm NL} $ to be
degenerate with other parameters and thus to have a larger  marginal error. 
Here we study the degeneracies among the large-scale  non-Gaussian
halo bias (for NG of the local type) and the cosmological parameters
which affect the large-scale halo power spectrum,
focusing on  dark energy perturbations, massive neutrinos,
number of relativistic species, and  running spectral index,
which can produce large deviations of the underlying cosmology from the
minimal $\Lambda$CDM scenario.
The paper is organized as follows. In \S \ref{Non-Gaussian halo bias}
we briefly review the analytic expressions of the non-Gaussian halo
power spectrum generalized to redshift dependent
transfer functions. The redshift dependence is due to the presence of
both dark energy perturbations and massive neutrinos.
In \S \ref{Methodology} we summarize the Fisher matrix formalism
applied to the observed halo power spectrum.
In \S \ref{Model parameters} we describe the assumed fiducial
cosmology. Finally, in \S \ref{Results} and \S \ref{Conclusions}
we discuss the results and draw our conclusions.
 
\section{Non-Gaussian halo bias}
\label{Non-Gaussian halo bias}
{Here we summarize the derivation of \cite{MV08}, extending it to the
  case of a redshift dependent transfer function}. In Fourier space, the filtered linear
over-density $\delta_R$ is related to the primordial potential
$\Phi({\bf k})$ by the Poisson equation:
\begin{eqnarray}
\delta_{R}({\bf k},z)=\frac{2}{3} \frac{D(z) T(k,z) k^2 c^2}
{H_0^2\Omega_{m,0}} W_R(k) \Phi({\bf k})
\equiv {\cal M}_R(k,z)\Phi({\bf k})\;,
\label{eq:defM}
\end{eqnarray}
where $T(k,z)$ denotes the matter transfer function (which is redshift
dependent in the presence of massive neutrinos and/or dark energy
perturbations), $W_R(k)$ is the Fourier transform of the top-hat function of width
$R$, $H_0$ and $\Omega_{m,0}$ are the current values of the Hubble constant
and the total matter energy density respectively, and
$D(z)=(1+z)^{-1}g(z)/g(0)$
is the linear growth-factor of density fluctuations,
normalized to $D(0)=1$ with $g(z)$ being the growth suppression factor
for non Einstein-de Sitter universes.

In this context, the non-Gaussian halo power-spectrum takes the form
\begin{eqnarray}
P_{\rm h}(k,z)=b_{L,h}^2(z,M) P_{\delta \delta}(k, z)
\left[ 1+4 f_{\rm NL} b_{L,h}(z,M) \beta(k,z) \right].
\label{eq:pkhalobias}
\end{eqnarray}
Here $b_{L,h}(z,M) \equiv q\delta_c(z)/(\sigma_M^2 D(z))$ is the
Gaussian lagrangian bias\footnote{In general, the Gaussian halo bias
may have a non trivial dependence on both the halo formation redshift
$z_f$ and the observation redshift $z_o$ \cite{Eetal88,CK89,MoWhite96}.
However, for objects that did not undergo recent mergers, $z_f \gg
z_o$, or in the case of rapid mergers $z_f \approx z_o$ for 
$\delta_c^2 \gg \sigma_M^2$, i.e. large $M$ and/or high $z_f$,                                     
the bias is well approximated by Eq.~(\ref{eq:pkhalobias}).}
for dark matter halos of mass $M$ \cite{Kaiser84},
$\sigma_M^2$ is the mass variance linearly extrapolated to $z=0$,
$\delta_c(z) \equiv \Delta_c(z)/D(z)$, being
$\Delta_c(z)$ the linear overdensity for spherical collapse,
which can be considered as a constant $\Delta_c(z)=1.686$, even
in the presence of dark energy \cite{Percival:0508156}, $q$ is a
factor extracted from N-body simulations (\cite{Grossietal09} and references therein)
and $P_{\delta \delta}(k, z)$ is the power spectrum of $\delta_R$ (as it
will be shown later in Eq.~(\ref {eq:Pm})).
Finally, $\beta(k,z)$ is  defined as
\begin{eqnarray}
\beta(k,z)\equiv\frac{1}{8\pi^2 {\cal M}_R(k,z)}\int dk_1 k_1^2 
{\cal M}_R(k_1,z) P_{\phi}(k_1)
\!\!\!\!\int_{-1}^1\!d\mu {\cal M}_R\left(\sqrt{\alpha},z\right)
\left[\!\frac{P_{\phi}\left(\sqrt{\alpha}\right)}{P_{\phi}(k)}+2\right]\!,\,\,\,
\label{alpha}
\end{eqnarray}
where $\alpha=k^2_1+k^2+2k_1k\mu$.
Here we adopt the so-called CMB $f_{\rm NL}$
normalization where Eq.~(\ref{eq:fnl}) is intended to be deep
in the matter-dominated era\footnote{See \cite{Grossietal09} for the large-scale structure-normalized conversions.}.
Consequently, the non-Gaussian halo Lagrangian bias reads\footnote{A
  more accurate expression is given by $b_{L,h}^{\rm NG}(z,M)=
  b_{L,h}(z,M)\sqrt(1+4 f_{\rm NL} b_{L,h}(z,M)\beta(k,z))$}
\begin{eqnarray}
\!b_{L,h}^{\rm NG}(z,M) \simeq b_{L,h}(z,M)[1+2 f_{\rm NL} b_{L,h}(z,M)\beta(k,z)] .
\label{bNG}
\end{eqnarray}
Making the standard assumption that halos move coherently
with dark matter, the Eulerian bias is $b_E=1+b_L$.
The halo power spectrum given by Eq.~(\ref{eq:pkhalobias}),
is connected directly to the underlying dark
matter power spectrum and can be recostructed from the galaxy power
spectrum using different techniques (e.g. \cite{Reidetal09:0907.1659}).
It provides important information on the growth of structure, which
helps in constraining dark energy and neutrino masses together
with primordial non-Gaussianities. We shall exploit information from
both the shape and the amplitude of the NG halo power spectrum
up to scales $k<0.03h$/Mpc, where $\beta(k,z)$ has a negligible
dependence on the halo mass (recall that for local non-Gaussianity
  $\beta(k,z) \propto 1/k^2$). In addition, Eqs.~(\ref{eq:pkhalobias})-(\ref{bNG}) are valid only on
  scales much larger than the Lagrangian radius of the halo
  \cite{MV08,CVM08} and have been tested and calibrated on N-body simulations only on scales
  $k<0.03h$/Mpc~\cite{Grossietal09}. We believe that  the approach followed here is 
conservative, since in our parameter forecasts we do not consider scales $k>0.03h$/Mpc in the halo power spectrum,
which could provide much tighter constraints on cosmology, via
e.g. Baryonic Acoustic Oscillations (BAO).
In what follows, we shall use Eq.~(\ref{eq:pkhalobias}) to propagate
errors of the non-Gaussian halo power spectrum into errors of the
cosmological parameters via a Fisher matrix approach, as described below.

\section{Methodology}
\label{Methodology}
In this paper we adopt the Fisher matrix formalism to make predictions
of the cosmological parameter errors including the NG parameter
$f_{\rm NL}$.
The Fisher matrix is defined as the
second derivative of the likelihood surface about the maximum. As long
as the posterior distribution for the parameters is well approximated
by a multivariate Gaussian function, its elements are given by
\cite{Tegmark,Jungman,Fisher}
\begin{equation}
\label{eq:fish}
F_{\alpha\beta}=\frac{1}{2}{\rm
  Tr}\left[C^{-1}C_{,\alpha}C^{-1}C_{,\beta}\right]~,
\end{equation}
where $C=S+N$ is the total covariance which consists of signal
$S$ and noise $N$ terms. The commas in Eq.~(\ref{eq:fish})
denote derivatives with respect to the cosmological parameters
within the assumed fiducial cosmology\footnote{In practice,
it can happen that the choice of parameterization                                         
makes the posterior distribution slightly non-Gaussian. However, even
in this case, the error introduced by assuming Gaussianity in the
posterior distribution can be considered as reasonably small, and therefore the
Fisher matrix approach still holds as an excellent approximation for
parameter forecasts.}.

To derive realistic parameter forecasts, we consider future
redshift surveys, as Euclid\footnote{http://sci.esa.int/euclid}- and
LSST\footnote{http://www.lsst.org/lsst/science/scientist\_dark\_energy}-like
galaxy surveys. While for BAO surveys
an accurate redshift measurement is crucial \cite{SeoEisenstein03, BlakeBridle05},
for our purposes, a precise redshift extraction is not needed, as it
will be explained after Eq.~(\ref{eqn:fisher}). For the EUCLID-like survey we will assume a redshift
coverage of $0.5<z<2$ and a sky area of $f_{\rm sky}=20000~$deg$^2$.
For the LSST-like survey,  we will consider $0.3<z<3.6$ and
$f_{\rm sky}=30000~$deg$^2$ for redshift and area coverages,
respectively.
In order to explore the cosmological parameters
constraints from a given redshift survey, it is mandatory to specify
the measurement uncertainties of the halo power spectrum.
In general, the statistical error on the measurement of $P_{\rm h}(k)$
at a given wavenumber bin is given by \cite{FKP}
\begin{equation}
\left[\frac{\Delta P_{\rm h}}{P_{\rm h}}\right]^2=
\frac{2(2\pi)^2 }{V k^2\Delta k\Delta \mu}
\left[1+\frac1{\bar{n}_{\rm h}P_{\rm h}}\right]^2,
\label{eqn:pkerror}
\end{equation}
where $\bar{n}_{\rm h}$ is the mean number density of dark matter
halos, $V$ is the comoving survey volume of the galaxy survey, and $\mu$
is the cosine of the angle between $\bf{k}$ and the line-of-sight
direction.

To our purposes it is adequate to perform an angular average over
$\mu$.  Thus, our Fisher matrix for the large-scale
structure data is given by
\begin{eqnarray}
F^{\rm LSS}_{\alpha\beta}=2\frac{V}{8\pi^2}\int^{k_{\rm max}}_{k_{\rm
    min}}k^2dk~
\frac{\partial \ln P_{\rm h}(k)}{\partial p_\alpha}
\frac{\partial \ln P_{\rm h}(k)}{\partial p_\beta}
 \left[\frac{\bar{n}_{\rm h}P_{\rm h}(k)}{\bar{n}_{\rm
      h}P_{\rm h}(k)+1}\right]^2,
\label{eqn:fisher}
\end{eqnarray}
where $p_\alpha$ represents the chosen set of cosmological parameters.

We divide the surveys in redshift bins of width $\Delta z=0.1$
(larger than standard photometric and spectroscopic redshift
errors), and set $k_{\rm max}$ to be $0.03h$/Mpc and $k_{\rm min}$
to be greater than $2\pi/\Delta V^{1/3}$, where $\Delta V$
is the volume of the redshift shell. Conservatively, we do not consider
here that scales larger than $k_{\rm min}$ can be used by cross-correlating
different shells.

The effect of NG alters the broad-band behavior of $P_{\rm h}(k)$
on very large scales, where $P_{\rm h}(k)$ is unaffected by the
precision
with which the radial positions of the galaxies are measured.
Thus, we can treat photometric and spectroscopic surveys on
the same footing. Moreover, 
the requirement of surveying a large volume
of the universe and sampling highly biased galaxies to beat
shot-noise, which is a key point for BAO surveys, is also a bonus
for constraining primordial NG \cite{CVM08}. 

Note, moreover, that the NG
correction of the halo bias 
is boosted by the Lagrangian Gaussian halo bias factor itself.

In particular, for the value of $k_{\rm min}$ used here, we find that
$P_{\rm h}(k_{\rm min})\simeq P_{\rm h}(k=0.2 h{\rm/Mpc})$, thus, the
shot-noise
requirement for BAO surveys of $\bar{n}P(k=0.2{\rm h/Mpc}) >  1$,
implies that for all scales of interest here, $\bar{n}P\gg 1$.
We have checked that our results do not change if we impose 
$\bar{n}P \sim 3$ at all scales.

We compute as well the CMB Fisher matrix to obtain
forecasts for the Planck satellite\footnote{www.rssd.esa.int/PLANCK}.
We follow here the method of \cite{Licia05}, considering 
the likelihood function for a realistic experiment with partial sky coverage,
and noisy data
{\small
\begin{eqnarray}
&&-2\ln{\cal L}=\sum_{\ell} (2\ell+1)\Bigg\{f_{sky}^{BB}\ln\left(\frac{\CC_{\ell}^{BB}}{\hat{\CC}_{\ell}^{BB}}\right)+
\sqrt{f_{sky}^{TT}f_{sky}^{EE}}\ln\left(\frac{\CC_{\ell}^{TT}\CC_{\ell}^{EE}-
(\CC_{\ell}^{TE})^2}{\hat{\CC}_{\ell}^{TT}\hat{\CC}_{\ell}^{EE}-
(\hat{\CC}_{\ell}^{TE})^2}\right) \nonumber\\
&&+\sqrt{f_{sky}^{TT}f_{sky}^{EE}}\frac{\hat{\CC}_{\ell}^{TT}\CC_{\ell}^{EE}+
\CC_{\ell}^{TT}\hat{\CC}_{\ell}^{EE}-
2\hat{\CC}_{\ell}^{TE}\CC_{\ell}^{TE}}{\CC_{\ell}^{TT}\CC_{\ell}^{EE}-
(\CC_{\ell}^{TE})^2}+f_{sky}^{BB}\frac{\hat{\CC}_{\ell}^{BB}}{\CC_{\ell}^{BB}}-2\sqrt{f_{sky}^{TT}f_{sky}^{EE}}-f_{sky}^{BB}\Bigg\}
\label{eq:like_real}
\end{eqnarray}}
and compute its second derivatives to obtain the corresponding Fisher matrix
\begin{equation}
F^{\rm CMB}_{\alpha\beta}=\left \langle -\frac{\partial^2 L}{\partial p_\alpha \partial p_\beta}
\right \rangle|_{{\bf p}=\bar{{\bf p}}} \, .
\label{standard_fish}
\end{equation}
In Eq.~(\ref{eq:like_real}) $\CC^{XY}_{\ell}=\Cc^{XY}_{\ell}+{\cal
N}^{XY}_{\ell}$ being $\Cc^{XY}_{\ell}$ the temperature and
polarization power spectra ($X,Y \equiv \{T,E,B\}$) and 
${\cal N}_{\ell}$ the noise bias. Finally, $f_{sky}^{XY}$ is the
fraction of observed sky which can be different for the $T$-, $E$-,
and $B$-modes.
In Eq.~(\ref{standard_fish}) $L\equiv \ln {\cal L}$, $p_\alpha$ and
$p_\beta$ denote the cosmological parameters of the assumed model and form the vector ${\bf p}$ whose
fiducial value is given by $\bar{{\bf p}}$.

Combining the Planck and redshift survey Fisher matrices
($F_{\alpha\beta} =F^{\rm LSS}_{\alpha\beta}+F^{\rm CMB}_{\alpha\beta}$) we get the
joint constraints for ${\bf p}$. The 1--$\sigma$ error on $p_\alpha$ marginalized over the
other parameters is
$\sigma(p_\alpha)=\sqrt{({F}^{-1})_{\alpha\alpha}}$, 
being ${F}^{-1}$ the inverse of the Fisher matrix.
We then consider joint constraints in a two-parameter
subspace (marginalized over all other cosmological parameters) to study the covariance between $f_{\rm NL}$ and the other cosmological parameters considered in this work.

Furthermore, in order to quantify the level of degeneracy between
different parameters and $f_{\rm NL}$, we estimate the so-called
correlation coefficients, given by
\begin{equation}
r\equiv \frac{({F}^{-1})_{p_{f_{\rm NL}}p_\alpha}}
{\sqrt{({{F}}^{-1})_{p_{f_{\rm NL}}p_{{f_{\rm
            NL}}}}{({F}^{-1})_{p_\alpha p_\alpha}}}},
\label{correlation}
\end{equation}
where $p_\alpha$ denotes one of the model parameters.  When the
coefficient $|r|=1$, the two parameters are totally degenerate, while
$r=0$ means they are uncorrelated.
\begin{table}[!ht]
\begin{center}
\caption{Correlation coefficients for $M_\nu$--cosmology}
\begin{tabular}{c c c c c c c c c c c}
\hline\hline
& &$\Omega_b h^2$ & $h$ & $\Omega_{c,0} h^2$ & $\Delta^2_{\cal
  R}(k_0)$ &$n_s$ & $w$ &$c_s^2$ &$\alpha_s$ &$M_\nu$\\
\hline
\footnotesize{LSST} &$f_{\rm NL}$& $0.50$ & $0.31$ & $0.36$ &
$0.35$ & $-0.03$ & $0.41$ & $-0.32$ & $-0.31$ & $0.03$
\\
\footnotesize{LSST+Planck} &$f_{\rm NL}$& $-0.22$ & $-0.19$ &
$0.36$ & $-0.12$ & $0.14$ & $-0.07$ & $0.10$ & $-0.13$ & $0.34$
\\
\\
\footnotesize{EUCLID} &$f_{\rm NL}$& $0.55$ & $0.25$ & $0.31$ & $0.36$ & $-0.06$ &
$0.39$ & $-0.30$ & $-0.24$ & $0.09$
\\
\footnotesize{EUCLID+Planck} &$f_{\rm NL}$& $-0.17$ & $-0.05$ & $0.24$ & $-0.07$ &
$0.06$ & $-0.07$ & $0.09$ & $-0.05$ & $0.16$\\
\hline
\label{correlation_mnu0.3}
\end{tabular}
\end{center}
\end{table}
\begin{table}[!h]
\begin{center}
\caption{Correlation coefficients for $N_\nu^{\rm rel}$--cosmology}
\begin{tabular}{c c c c c c c c c c c}
\hline\hline
& & $\Omega_{b,0} h^2$ & $h$ & $\Omega_{c,0} h^2$ &
$\Delta^2_{\cal R}(k_0)$ & $n_s$ &$w$ &$c_s^2$ &$\alpha_s$
&$N_\nu^{\rm rel}$\\
\hline
\footnotesize{LSST} &$f_{\rm NL}$& $-0.06$ & $0.34$ & $0.12$ &
$-0.13$ & $0.12$ & $0.34$ & $-0.39$ & $-0.44$ & $-0.64$
\\
\footnotesize{LSST+Planck} &$f_{\rm NL}$& $0.24$ & $-0.48$ & $0.61$ &
$-0.24$ & $0.19$ & $0.54$ & $0.04$ & $-0.08$ & $0.69$
\\
\\
\footnotesize{EUCLID} &$f_{\rm NL}$& $-0.05$ & $0.34$ &
$0.08$ & $-0.20$ & $0.05$ & $0.11$ & $-0.26$ & $-0.45$ & $-0.69$
\\
\footnotesize{EUCLID+Planck} &$f_{\rm NL}$& $0.34$ & $-0.57$ &
$0.76$ & $-0.26$ & $0.21$ & $0.64$ & $0.08$ & $-0.05$ & $0.77$\\
\hline
\label{correlation_Nnu}
\end{tabular}
\end{center}
\end{table}
\begin{table}[!ht]
\begin{center}
\caption{$f_{\rm NL}$ 1-$\sigma$ errors for $M_\nu$--cosmology}
\begin{tabular}{c c c c c}
\hline\hline
fixed parameter &LSST & LSST+PLANCK & EUCLID & EUCLID+PLANCK\\
\hline
non-marginalized & $1.65$ & $1.65$ & $2.79$ & $2.79$
\\
marginalized & $4.52$ & $2.11$ & $8.86$ & $3.12$
\\
$\Omega_{b,0}h^2$ & $3.91$ & $2.06$ & $7.41$ & $3.07$ 
\\
$h$ & $4.29$ & $2.07$ & $8.58$ & $3.11$
\\
$\Omega_{c,0}h^2$ & $4.22$ & $1.97$ & $8.43$ & $3.02$
\\
$\Delta^2_{\cal R}(k_0)$ & $4.22$ & $2.10$ & $8.27$ & $3.11$
\\
$n_s$ & $4.52$ & $2.09$ & $8.85$ & $3.11$
\\
$w$ & $4.12$ & $2.11$ & $8.16$ & $3.11$
\\
$c_s^2$ & $4.27$ & $2.10$ & $8.44$ & $3.11$
\\
$\alpha_s$ & $4.30$ & $2.10$ & $8.61$ & $3.11$
\\
$M_\nu$ & $4.52$ & $1.99$ & $8.83$ & $3.08$
\\
$\alpha_s, c_s^2$ & $4.05$ & $2.09$ & $8.13$ & $3.10$
\\
$\alpha_s, M_\nu$& $4.28$ & $1.94$ & $8.57$ & $3.07$
\\
$c_s^2, M_\nu$ & $4.26$ & $1.97$ & $8.35$ & $3.06$
\\
$c_s^2, M_\nu, \alpha_s$ & $4.02$ & $1.93$ & $8.03$ & $3.06$\\
\hline
\label{fnl_errors_mnu0.3}
\end{tabular}
\end{center}
\end{table}
\begin{table}[!h]
\begin{center}
\caption{$f_{\rm NL}$ 1-$\sigma$ errors for $N_\nu^{\rm rel}$--cosmology}
\begin{tabular}{c c c c c}
\hline\hline
fixed parameter & LSST & LSST+PLANCK & EUCLID & EUCLID+PLANCK\\
\hline
non-marginalized & $1.46$ & $1.46$ & $2.56$ & $2.56$ 
\\
marginalized & $5.08$ & $2.56$ & $10.15$ & $4.79$
\\
$\Omega_{b,0}h^2$ & $5.07$ & $2.49$ & $10.13$ & $4.51$
\\ 
$h$ & $4.77$ & $2.25$ & $9.53$ & $3.93$ 
\\
$\Omega_{c,0}h^2$ & $5.05$ & $2.03$ & $10.12$ & $3.13$
\\
$\Delta^2_{\cal R}(k_0)$ & $5.04$ & $2.49$ & $9.94$ & $4.62$
\\
$n_s$ & $5.05$ & $2.51$ & $10.13$ & $4.68$
\\
$w$ & $4.78$ & $2.16$ & $10.09$ & $3.67$
\\
$c_s^2$ & $4.69$ & $2.56$ & $9.79$ & $4.77$
\\
$\alpha_s$ & $4.57$ & $2.55$ & $9.06$ & $4.78$
\\
$N_\nu^{\rm rel}$ & $3.88$ & $1.86$ & $7.32$ & $3.06$
\\
$\alpha_s, c_s^2$ & $4.16$ & $2.55$ & $8.70$ & $4.77$
\\
$\alpha_s, N_\nu^{\rm rel}$ & $3.76$ & $1.76$ & $7.15$ & $2.92$
\\
$c_s^2, N_\nu^{\rm rel}$ & $3.57$ & $1.86$ & $7.15$ & $3.05$
\\
$c_s^2, N_\nu^{\rm rel}, \alpha_s$ & $3.43$ & $1.76$ & $6.95$ & $2.91$\\
\hline
\label{fnl_errors_Neff}
\end{tabular}
\end{center}
\end{table}
\section{Model parameters}
\label{Model parameters}
The Fisher matrix approach propagates errors of the observed 
$P_{\rm  h}$, see Eq.~(\ref{eqn:pkerror}), into errors of the cosmological
parameters which characterize the underlying fiducial cosmology.
According to the latest observations (e.g. \cite{Komatsuetal2010} and
refs. therein), our fiducial $\Lambda$CDM cosmological parameters are:
$\Omega_{m,0} h^2=0.1358$, $\Omega_{b,0} h^2=0.02267$,
$h=0.705$,  $\Delta^2_{\cal R}(k_0)=2.64\times 10^{-9}$, 
$n_s=0.96$, $\alpha_s=0$ and $f_{\rm NL}=0$. Here $\Omega_{m,0}$ and
$\Omega_{b,0}$ are the total matter and baryon present-day energy densities, 
respectively,
in units of the critical energy density of the Universe, $h$ is given
by $H_0=100 h$ km s${}^{-1}$ Mpc${}^{-1}$ , where $H_0$ is the Hubble
constant, $\Delta^2_{\cal R}(k_0)$ represents
the dimensionless amplitude of the primordial curvature perturbations
evaluated at a pivot scale $k_0$, $n_s$ is the scalar spectral index
of the primordial matter power spectrum, assumed to be a power-law,
and $\alpha_s$ is the running of the scalar spectral index.
We will consider two different fiducial models matching this $\Lambda$CDM cosmology  as follows.
We adopt  the same values for the 7 ``base" parameters.
We do not consider primordial gravitational waves
and assume a flatness prior, $\Omega_{\rm K}=0$, as predicted
by long-lasting inflation models,
so that $\Omega_{de,0}=1-\Omega_{m,0}$, where $\Omega_{\rm K}$ and
$\Omega_{de,0}$ are, respectively, the present-day 
energy densities associated to the spatial curvature and
to the dark energy component of the Universe, in units of the critical density.

A Gaussian prior of $5$\% on the present-day Hubble's
constant $H_0=100 h~{\rm km~s^{-1}~Mpc^{-1}}$ is assumed,
following the results of \cite{Riess et al. (2009) 0905.0695}.
While this uncertainty is comparable to the one achieved by recent 
WMAP-7yr data\footnote{http://lambda.gsfc.nasa.gov/} in the
determination of $H_0$ for the $\Lambda$CDM model~\cite{Komatsuetal2010}, 
this information will
improve, as expected, the parameter constraints on models different from 
the minimal $\Lambda$CDM model, 
such as models with
dark energy perturbations, massive neutrinos and non-vanishing running
of the spectral index.

We also consider dark energy to be a cosmic
fluid with clustering properties on the Gpc scale, described by an
equation of state that we assume to be constant
\begin{equation}
w\equiv \frac{p_{de}}{\rho_{de}}=w|_{\rm fid}\; ,
\end{equation}
where where $p_{de}$ and $\rho_{de}$ represent
  respectively the pressure and               
energy density of the dark energy fluid, and we assume
a fiducial value $w|_{\rm fid}=-0.9$ which lies
well within the current $95$\% C.L. limits on a constant dark
energy equation of state parameter $w$. The dark energy fluid
will be also described here by an effective sound speed $c_s$
which parametrizes the transition between the smooth and clustered
dark-energy regimes \cite{9801234}
\begin{equation}
c_s^2\equiv\frac{\delta p_{de}}{\delta\rho_{de}}~,
\end{equation}
whit fiducial value $c_s^2|_{\rm fid}=0.9$.
Dark energy perturbations will arise only if
the dark energy equation of state parameter $w$ is different from
$-1$.

We assume the power spectrum of primordial curvature
perturbations,
$P_{\cal R}(k)$, to be
\begin{equation}
\Delta^2_{\cal R}(k) \equiv \frac{k^3P_{\cal R}(k)}{2\pi^2}
= \Delta^2_{\cal R}(k_0)\left(\frac{k}{k_0}\right)^{n_s-1+\frac{1}{2}
\alpha_s \ln(k/k_0)}.             
\label{eq:pR}
\end{equation}
where $k_0=0.002$/Mpc and $\Delta^2_{\cal R}(k_0)|_{\rm
  fid}=2.64\times 10^{-9}$ \cite{arXiv:1001.4635}.

The matter energy density $\Omega_{m,0}$ includes the neutrino
contribution when neutrinos are non-relativistic
\begin{equation}
\Omega_{m,0} = \Omega_{c,0} + \Omega_{b,0} + \Omega_{\nu,0}~,
\label{Omega_m}
\end{equation}
where $\Omega_{\nu,0}$ is related to the sum of neutrino masses
$M_\nu\equiv\sum m_\nu$ as
\begin{equation}
\Omega_{\nu,0} = \frac{M_\nu}{93.8h^2~{\rm eV}}~,
\end{equation}
and the neutrino mass eigenstates are assumed to have a
{\it degenerate} spectrum, i.e. the three neutrinos have
the same mass.


We will constrain the following set of ``baseline" parameters
\begin{equation}
p_\alpha=\left\{\Omega_{b,0} h^2, h, \Omega_{c,0} h^2, \Delta^2_{\cal
    R}(k_0), 
n_s, w, c_s, \alpha_s, f_{\rm NL}\right\}\;.                                                        
\label{p_set}
\end{equation}
From here we specify two cosmological  models:
\begin{itemize} 
\item $N_\nu^{\rm rel}$--cosmology, where neutrinos are effectively massless  but the
the number of relativistic species $N_\nu^{\rm rel}$ can deviate from the standard value $N_\nu^{\rm rel}=3.04$. In this case the fiducial value $N_\nu^{\rm rel}|_{\rm fid}=3.04$ is chosen, fixing
$M_{\nu}={\rm const}=0$.
$N_\nu^{\rm rel}$ is given by the energy density associated to radiation
\begin{equation}
\Omega_{r,0} = \Omega_{\gamma, 0} \left(1+0.2271 N_\nu^{\rm rel}\right)~,
\end{equation}
where $\Omega_{\gamma, 0}=2.469\times 10^{-5} h^{-2}$ is the
present-day
photon energy density parameter for $T_{\rm cmb}=2.725$~K
\cite{Komatsuetal08}. 

\item $M_\nu$--cosmology, where $N_\nu^{\rm rel}$ is fixed at the
  fiducial value and  $M_\nu$ is allowed to vary. In this case, being
  still 
consistent with current data \cite{0910.0008,Komatsuetal2010} we
choose a fiducial value $M_\nu|_{\rm fid}=0.3$ eV. 
This choice is motivated by the fact that for taking two-sided
numerical derivatives, 
the fiducial $M_\nu$ must be non-zero. It is well known that the error
from cosmological 
observations on the neutrino mass depends somewhat on the fiducial
mass chosen; 
from  \cite{Kitchingnu} we estimate that around a fiducial  
$M_\nu=0$ the error on  $M_\nu=0$ would increase by less than 20\%. 
As it will be clear from Sec.~\ref{Results} the effect of this
correction  
will be negligible on the $f_{\rm NL}$ error estimate. 
In addition most of the signal to constrain $M_\nu$ from LSS surveys
will come 
from smaller scales (not considered here).
\end{itemize}

At the CMB level, if neutrinos are still relativistic
at the decoupling epoch, $z \simeq 1090$, i.e. if the
mass of the heaviest neutrino specie is $m_\nu<0.58$~eV,
massive neutrinos do not affect the CMB power spectra, except through
the gravitational lensing effect \cite{Komatsuetal08}, and, as a
consequence, the dark energy equation of state $w$ is not degenerate with the
neutrino mass. However, the limit on the the sum of the neutrino
masses degrades significantly when the dark energy
equation of state is a function of redshift, if the amplitude
of the galaxy spectrum is used for getting constraints on $w$ and $M_\nu$,
since dark energy and massive neutrinos both affect the growth rate of structures.

In this work, both massive neutrinos and
clustering properties of the dark energy perturbations are considered.
In this scenario,
the growth function of the dark matter perturbations is
scale-dependent, even at the linear level.
The overall effect induces a redshift-dependent
transfer function \cite{0512374,0606533,9710252}, and
the power spectrum of the linear density field, smoothed on a sphere of
radius $R$, takes the form
\begin{eqnarray}
P_{\delta\delta}(k,z)=\frac{8\pi^2c^4k_0\Delta^2_{\cal R}(k_0)}{25
  H_0^4\Omega_{m,0}^2}
W^2_R(k)T^2(k,z) D^2(z) 
\left(\frac{k}{k_0}\right)^{n_s+\frac{1}{2} \alpha_s \ln(k/k_0)}~,
\label{eq:Pm}
\end{eqnarray}
where $D(z)$ is the \emph{scale independent} linear
growth-factor defined in \S \ref{Non-Gaussian halo bias}. The redshift-dependent
transfer function  $T(k,z)$ is directly extracted from
CAMB\footnote{http://camb.info/}\cite{CAMB} 
at each redshift $z$, in order to compute the Fisher matrix,
given by Eq.~(\ref{eqn:fisher}), within each redshift bin.

Our analysis exploit exclusively the linear matter power spectrum,
since we restrict ourselves to scales $k \leqslant 0.03h$/Mpc,
 where the details of the  halo
occupation distribution of galaxies are irrelevant.
We focus here on dark matter halos with mass $\sim$
$10^{12}$--$10^{13}$ 
(where the lower mass limit is relative to the highest redshift of the survey).

Notice as well that in our Fisher matrix analysis we do not add
constraints on the $f_{\rm NL}$ parameter from CMB experiments,
so that our forecasts on $f_{\rm NL}$ result exclusively from
future redshift survey measurements of the dark matter halo power
spectrum on scales $k \leqslant 0.03 h$/Mpc, i.e. without including information from BAOs
which will further reduce the forecasted errors and residual degeneracies.
For all the reasons explained above, the results presented
in the next section should be considered conservative.

\section{Results}
\label{Results}
In this Section we present the predicted 1--$\sigma$ marginalized
errors of the
$f_{\rm NL}$ parameter, and the $f_{\rm NL}$ covariance with
the remaining cosmological parameters considered in our Fisher matrix
analysis. We show forecasts both from LSST and EUCLID data only,
as well as the expected errors after combining the results from these
two experiments with Planck forecasted errors.

First of all, when considering forecasts from the redshift surveys
alone, we expect that the $f_{\rm NL}$ parameter will be correlated
with all cosmological parameters which affect the amplitude and shape
of $P_{\rm h}(k)$ at scales $k \leqslant 0.03 h$/Mpc.
Tables~\ref{correlation_mnu0.3}-\ref{correlation_Nnu}
show the correlation $r$ (see Eq.~(\ref{correlation}))
among $f_{\rm NL}$ and the other cosmological parameters $p_\alpha$.
\begin{figure}
\begin{center}
\includegraphics[width=0.7\textwidth]{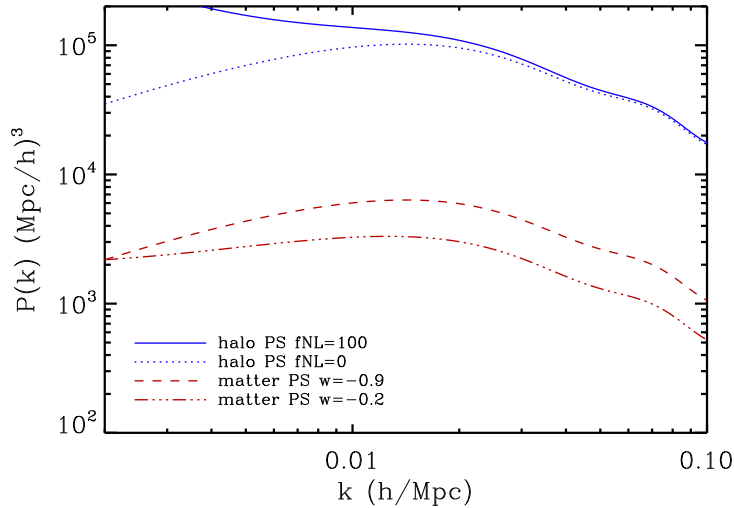}
\caption{The solid blue line represents the NG halo power spectrum for
$f_{\rm NL}=100$ and halo mass $M=10^{13} M_\odot$, while the dotted
blue line is the corresponding Gaussian halo power spectrum ($f_{\rm
  NL}=0$), at redshift $z=0$. 
The dashed red line represents the matter power spectrum
calculated for $w=-0.9$, while the red three-dot-dashed line is the
matter power spectrum for $w=-0.2$. The matter power spectra are evaluated at
$z=0$ and normalized to the same amplitude at $k=0.002$.
Note that $\Delta (\ln P_h(k))/\Delta w$ and $\Delta (\ln P_h(k))/\Delta
f_{\rm NL}$ have opposite sign in the range of $k$ of interest here:
it is clear from Eq. (3.3) that the two parameters can
compensate each other, i.e. they are correlated}
\label{weffect}
\end{center}
\end{figure}
We expect $w$ and $f_{\rm NL}$ to be correlated.
In fact, at scales $k \leqslant 0.03 h$/Mpc, an increase of $w$
produces a trend on the matter power
spectrum which is \emph{opposite} to the one produced in $P_{\rm
  h}(k)$ 
by increasing $f_{\rm NL}$. So the effect of
positively increasing $f_{\rm NL}$ can be mimicked by a larger $w$
(see Fig.~\ref{weffect}).

Likewise, $f_{\rm NL}$ is correlated with
$\Omega_{b,0} h^2$, $\Omega_{c,0} h^2$, and $M_{\nu}$, since, on the scales considered,  the
larger these
parameters are, the smaller the matter power spectrum is, and this competes with
the rise of $P_{\rm h}(k)$ due to an increasing value of $f_{\rm NL}$.

On the other hand, $f_{\rm NL}$ is negatively correlated with both
the running of the scalar spectral index $\alpha_s$ and the effective
sound speed of dark energy perturbations $c_s$. In fact, if either $\alpha_s$
or $c_s$ increase, the matter power spectrum $P_{\delta\delta}$
is modified  in a very similar way to the  halo power
spectrum $P_{\rm h}(k)$  for a larger $f_{\rm  NL}$ value.
Consequently, the effect of a positively increasing $f_{\rm NL}$ can
be mimicked by decreasing either $\alpha_s$ or $c_s$.
For what concerns the effective number of relativistic species, the
$f_{\rm NL}$--$N_\nu^{\rm rel}$  correlation is more complicated, as it
depends
on several factors. Naively, one would expect a positively correlation
between
these two parameters, since an increase in the number of
relativistic
particles should suppress the matter power spectrum, which
can be compensated by increasing $f_{\rm NL}$. However, since the
redshift evolution of the halo bias is also modified in this situation, the
correlation coefficient may change its sign.
\newline

When the Planck Fisher matrix information is added to the survey
Fisher matrix,  all degeneracies are either  resolved or drastically reduced. In some cases, the
correlation
coefficient $r$ can even change  sign, see
Tabs.~\ref{correlation_mnu0.3}-\ref{correlation_Nnu}.
This change in the behavior of $r$ arises either
due to the presence of dominant parameter degeneracies affecting
the CMB spectrum, or  because of 
marginalization of  a high-dimension parameter space
down to two variables. In particular, it is worth noting here that,
while
$f_{\rm NL}$ and $N_\nu^{rel}$ are negatively correlated if only
galaxy survey
data are considered,  they are positively correlated after adding
Planck priors. In fact, since both $N_\nu^{rel}$ and $\Omega_{c,0}h^2$ 
are strongly, positively correlated at the CMB level via the
equality redshift $z_{\rm eq}$, then, a positive correlation between
$\Omega_{c,0} h^2$ and $f_{\rm NL}$ automatically turns into a correlation of the
same sign
between $N_\nu^{\rm rel}$  and $f_{\rm NL}$.  Despite the residual
non-zero correlation coefficients, 
one should bear in mind that  the marginalized  $f_{\rm NL}$ errors decrease by a factor $>2$ when the CMB prior is added.

From our analysis, we conclude that the effective number of relativistic species is the main
parameter affecting the constraints on $f_{\rm NL}$.

The 1--$\sigma$ errors of $f_{\rm NL}$, for the two fiducial
cosmologies considered here, are shown in
Tabs.~\ref{fnl_errors_mnu0.3} and \ref{fnl_errors_Neff}. Let us notice
that the marginalized errors are significantly larger than the
non-marginalized ones when only the LSS surveys are used for the
forecasts. Nonetheless, the marginalized errors become comparable
in magnitude to the non-marginalized errors when Planck priors are
added, since the CMB mitigates the intrinsic degeneracies between
$f_{\rm NL}$ and the other cosmological parameters at the LSS 
level. It is also worth noting that the non-marginalized 1--$\sigma$
errors of $f_{\rm NL}$ presented in this work are larger 
than the corresponding errors presented in \cite{CVM08}. 
In fact we now consider less highly biased halos, with a fiducial bias
parameter more in-line with the expected one for (blue) EUCLID
galaxies \cite{Geachetal09,Orsietal09}.
It is important to note that the $f_{\rm NL}$ effect on the halo bias
is modulated by 
$b_{L,h}(z,M)$, which depends 
crucially on the selected halo and its merging history \cite{Slosar08,Reidinprep}.

Tabs.~\ref{fnl_errors_mnu0.3}-\ref{fnl_errors_Neff} show as well
the effect of each cosmological parameter on the $f_{\rm NL}$
forecasts. Obviously the parameters which have a larger impact on
the $f_{\rm NL}$ errors are the ones more degenerated with it, and
can be directly inferred from Tabs.~\ref{correlation_mnu0.3}-\ref{correlation_Nnu}.
Moreover, since with the inclusion of the parameters $\alpha_s$,
$c_s$, $M_\nu$ and $N_\nu^{rel}$, we have considered cosmologies
which deviate substantially from the minimal $\Lambda$CDM model, 
we fix pairs/triplets composed by these parameters
to show how much deviations from a $\Lambda$CDM cosmology can affect
the $f_{\rm NL}$ constraints.
For the $N_\nu^{\rm rel}$ model cosmology there is an important impact
on the $f_{\rm NL}$ marginalized errors from $\Omega_{c,0} h^2$,
  $h$, $w$ and, in particular, from $N_\nu^{\rm rel}$.
In summary, if $N_\nu^{\rm rel}$ is assumed to be fixed, the uncertainties on 
the other cosmological parameters increase the error on $f_{\rm NL}$
only by 10 to 30\%, depending on the survey.  If $N_\nu^{\rm rel}$ is
considered as an extra parameter to be  
simultaneously constrained from the data  
then the uncertainty in the underlying cosmology increases the $f_{\rm NL}$  error by $\sim 80\%$.

\begin{figure}[]
\begin{tabular}{c c c c} 
\includegraphics[width=3.5cm]{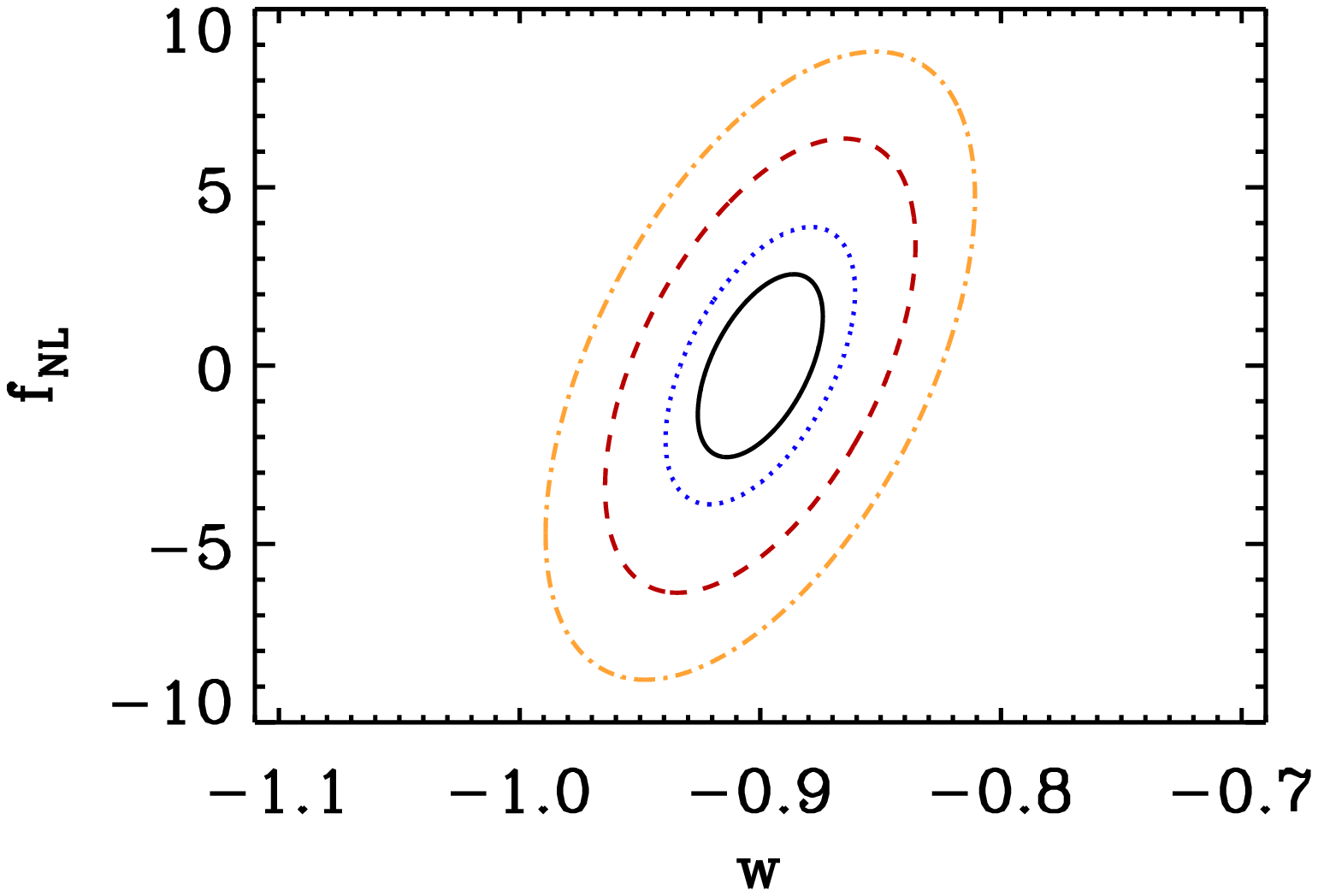}&
\includegraphics[width=3.5cm]{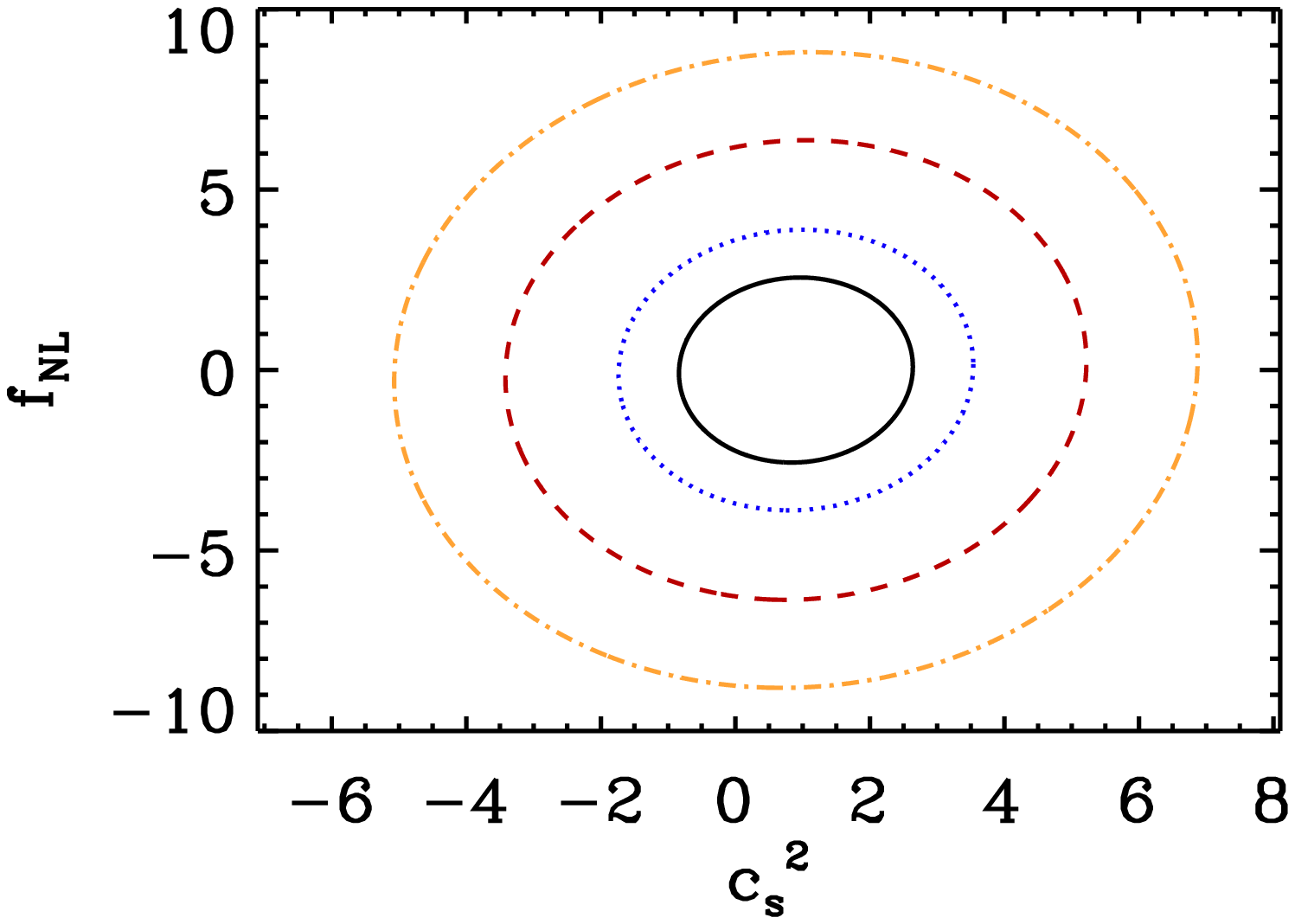}&
\includegraphics[width=3.5cm]{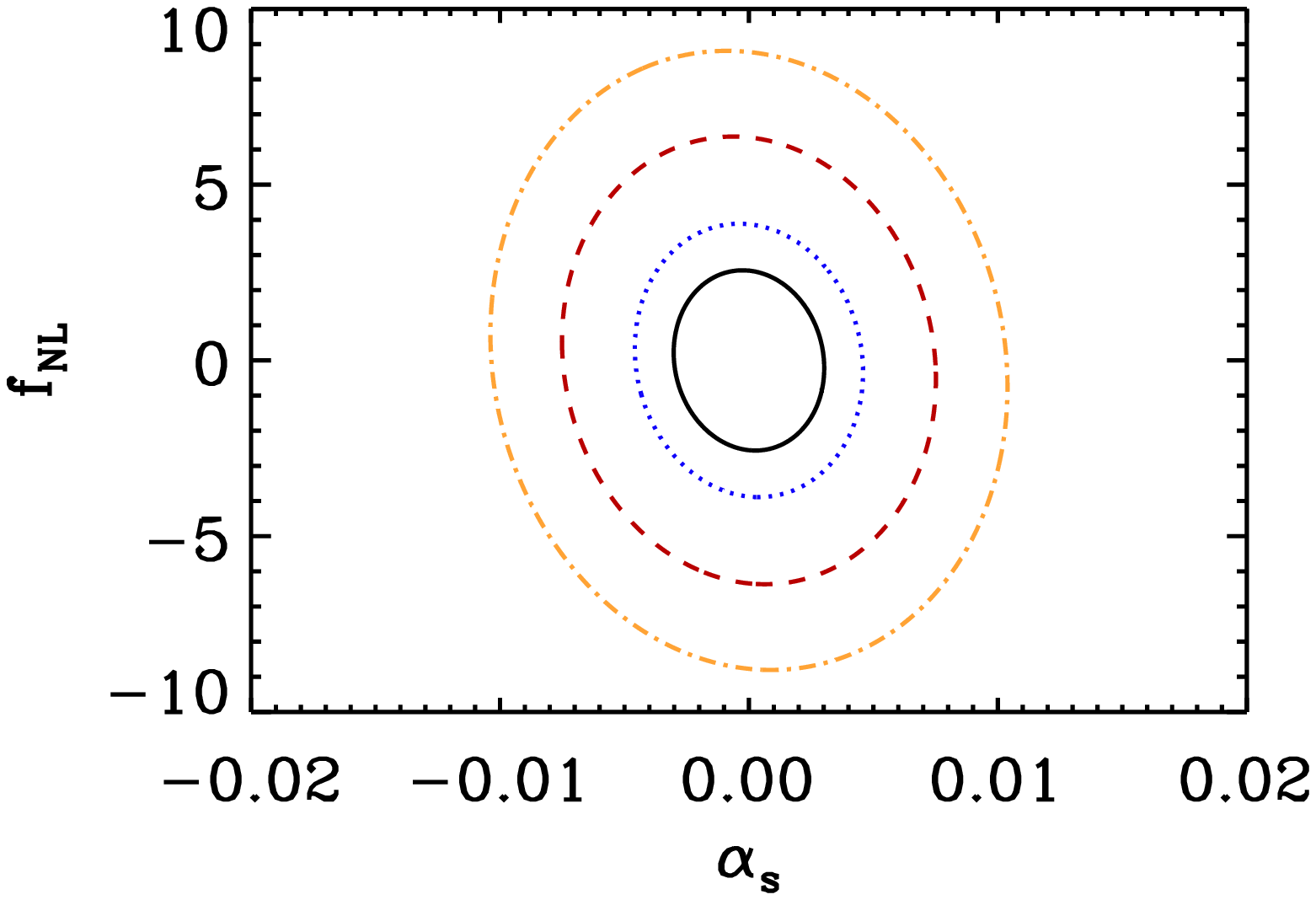}&
\includegraphics[width=3.5cm]{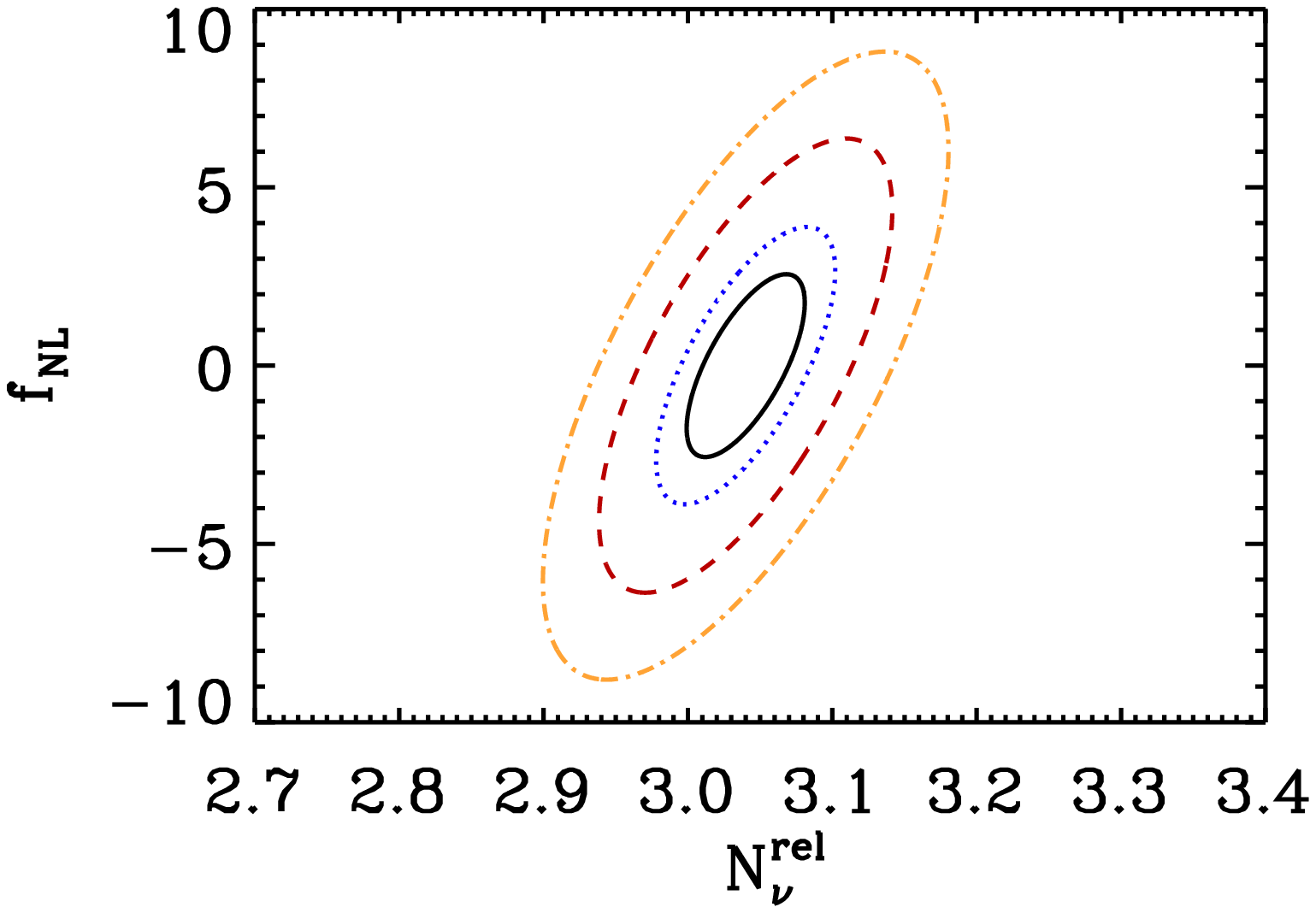}\\
\includegraphics[width=3.5cm]{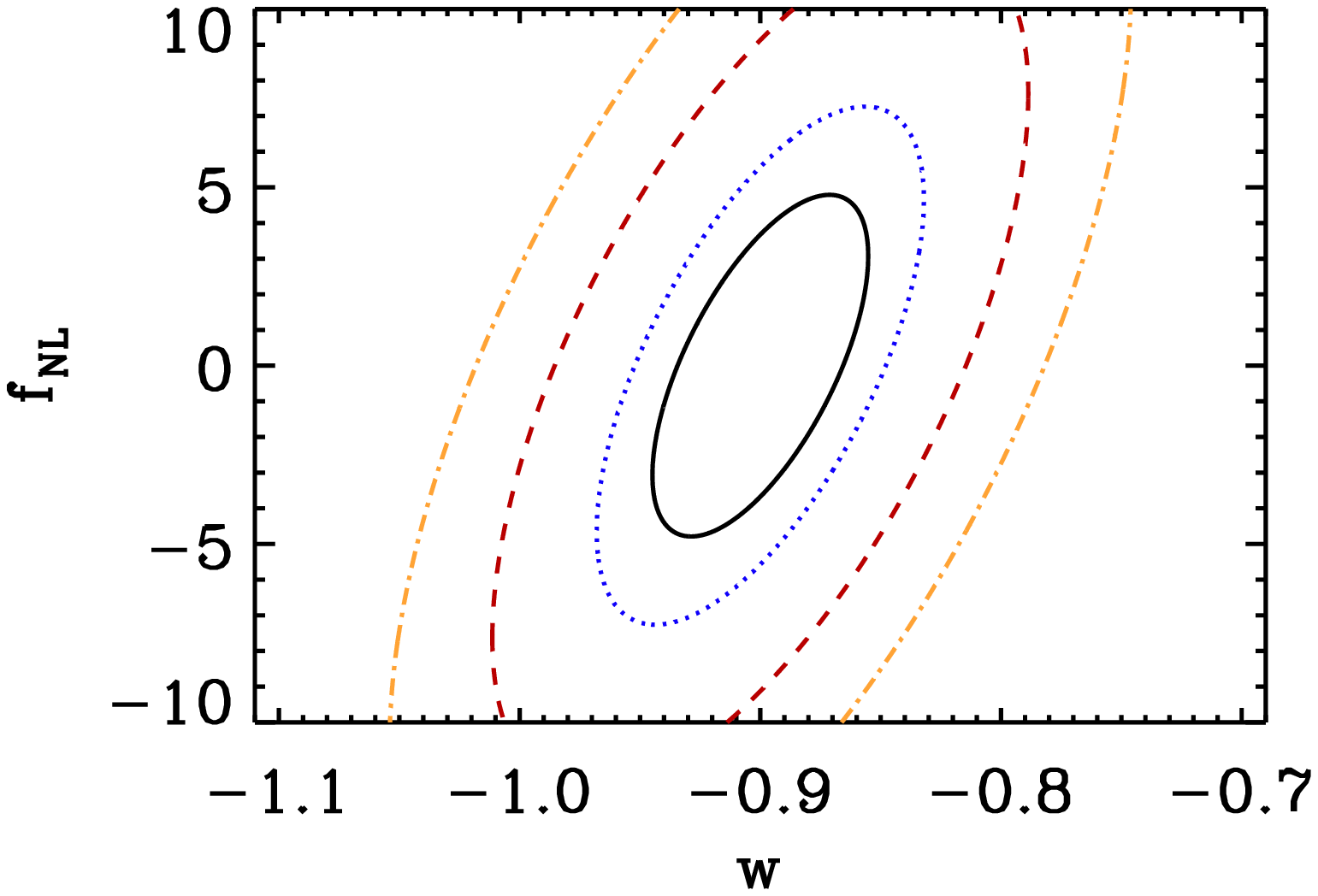}&
\includegraphics[width=3.5cm]{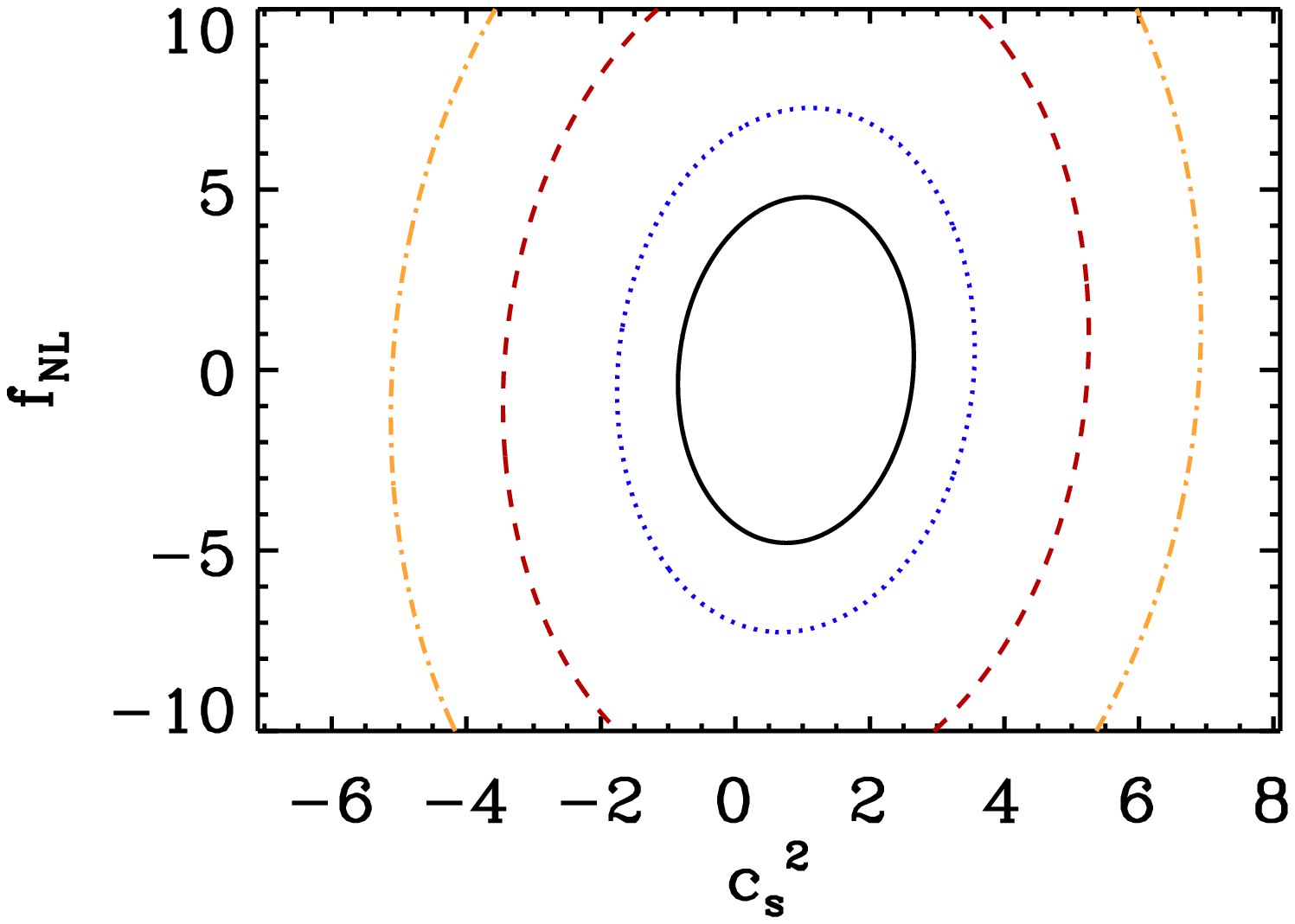}&
\includegraphics[width=3.5cm]{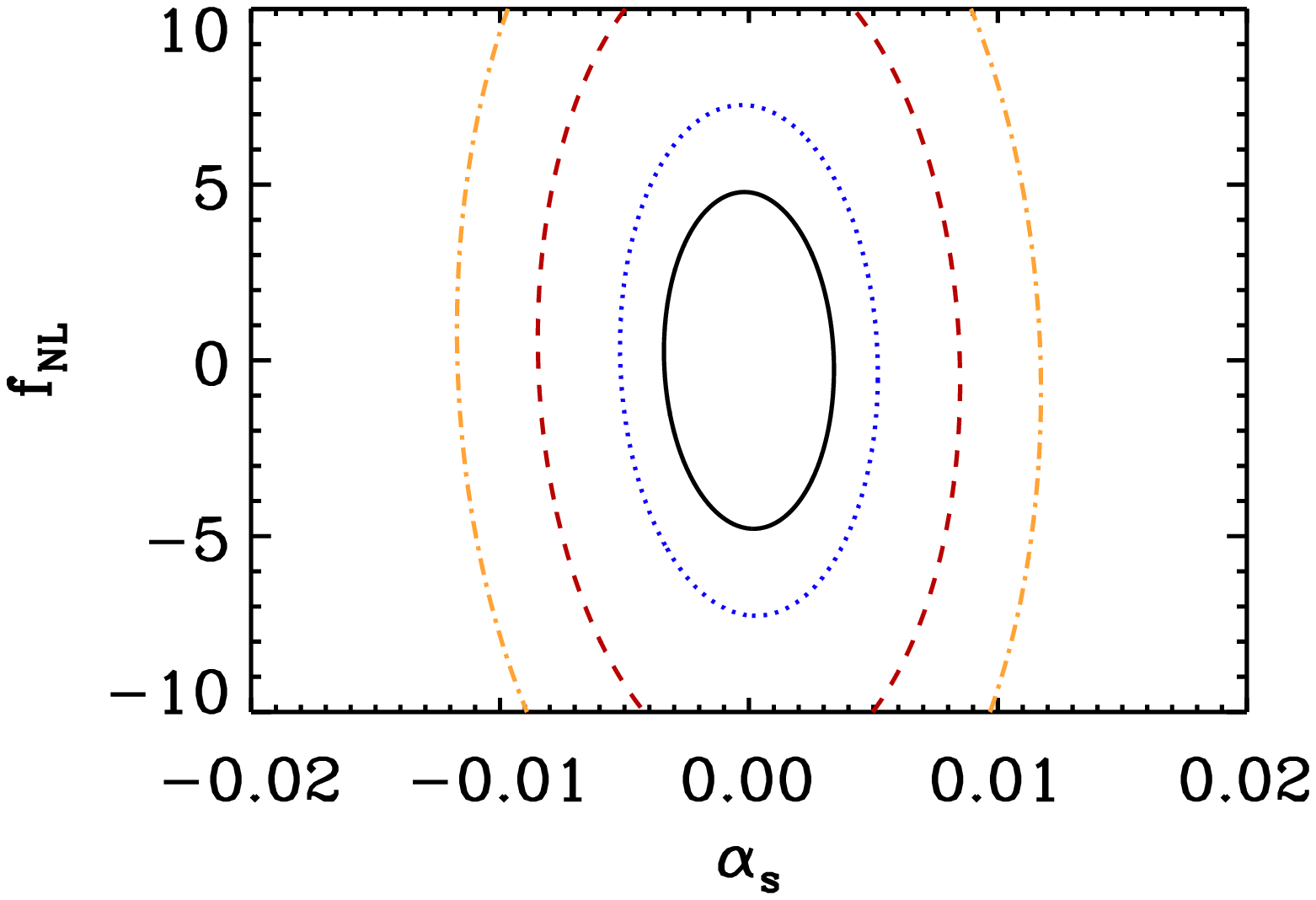}&
\includegraphics[width=3.5cm]{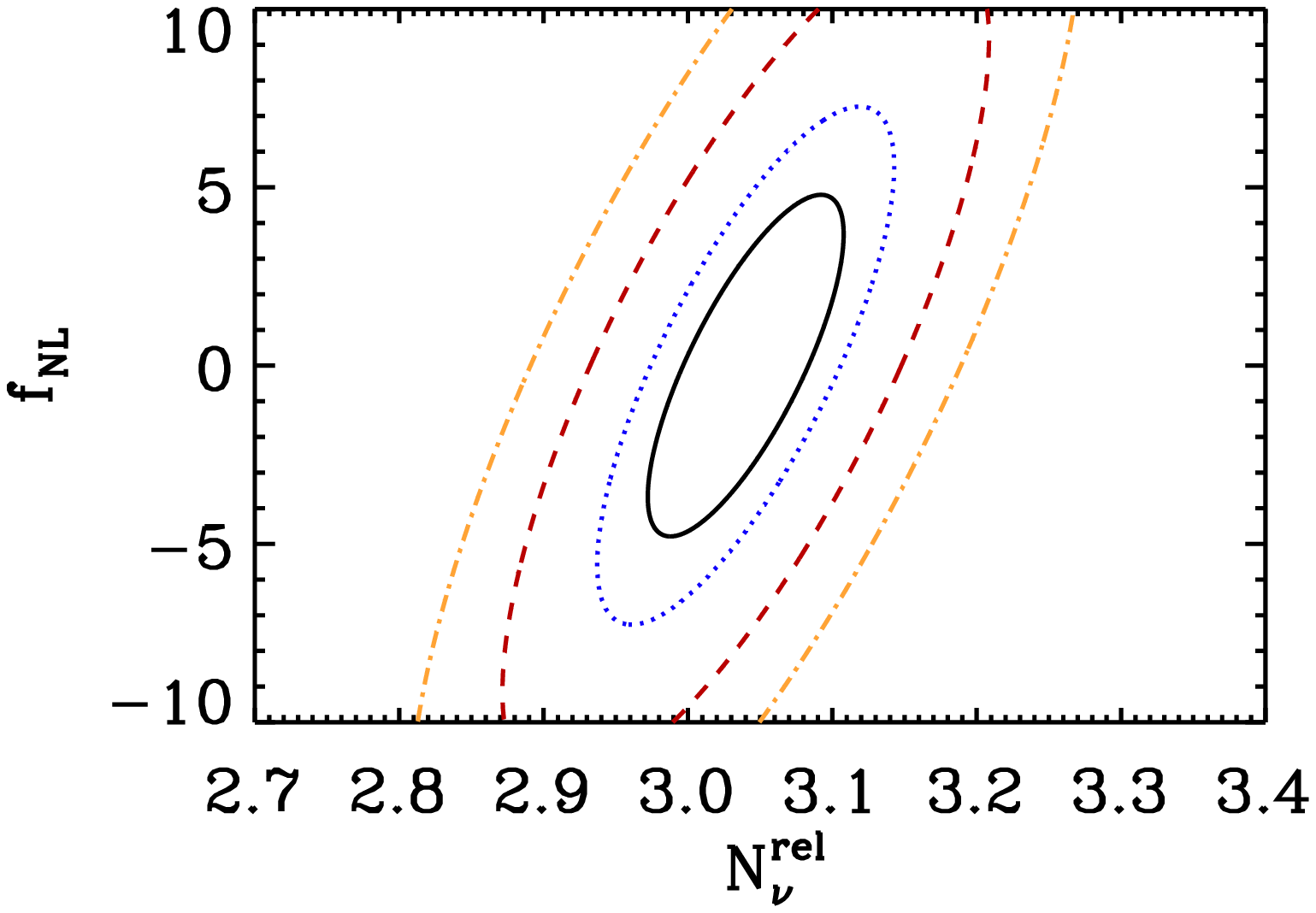}
 \end{tabular} 
\caption{2-parameter $f_{\rm NL}$-$p_\alpha$ joint contours for
  the fiducial model with extra relativistic degrees of freedom
  $N_\nu^{\rm rel}$ as described in the text, obtained after combining
  LSST (upper panels) and EUCLID (lower panels) data with Planck
  priors. The blue dotted line, the red dashed line and the orange
  dot-dashed line represent the 68$\%$ C.L., 95.4$\%$ C.L. and
  99.73$\%$ C.L., respectively. The black solid line shows the
  1-parameter confidence level at 1--$\sigma$.}
\label{fig_Neff}
\end{figure}

\begin{figure}[!h]
\begin{tabular}{c c c c} 
\includegraphics[width=3.5cm]{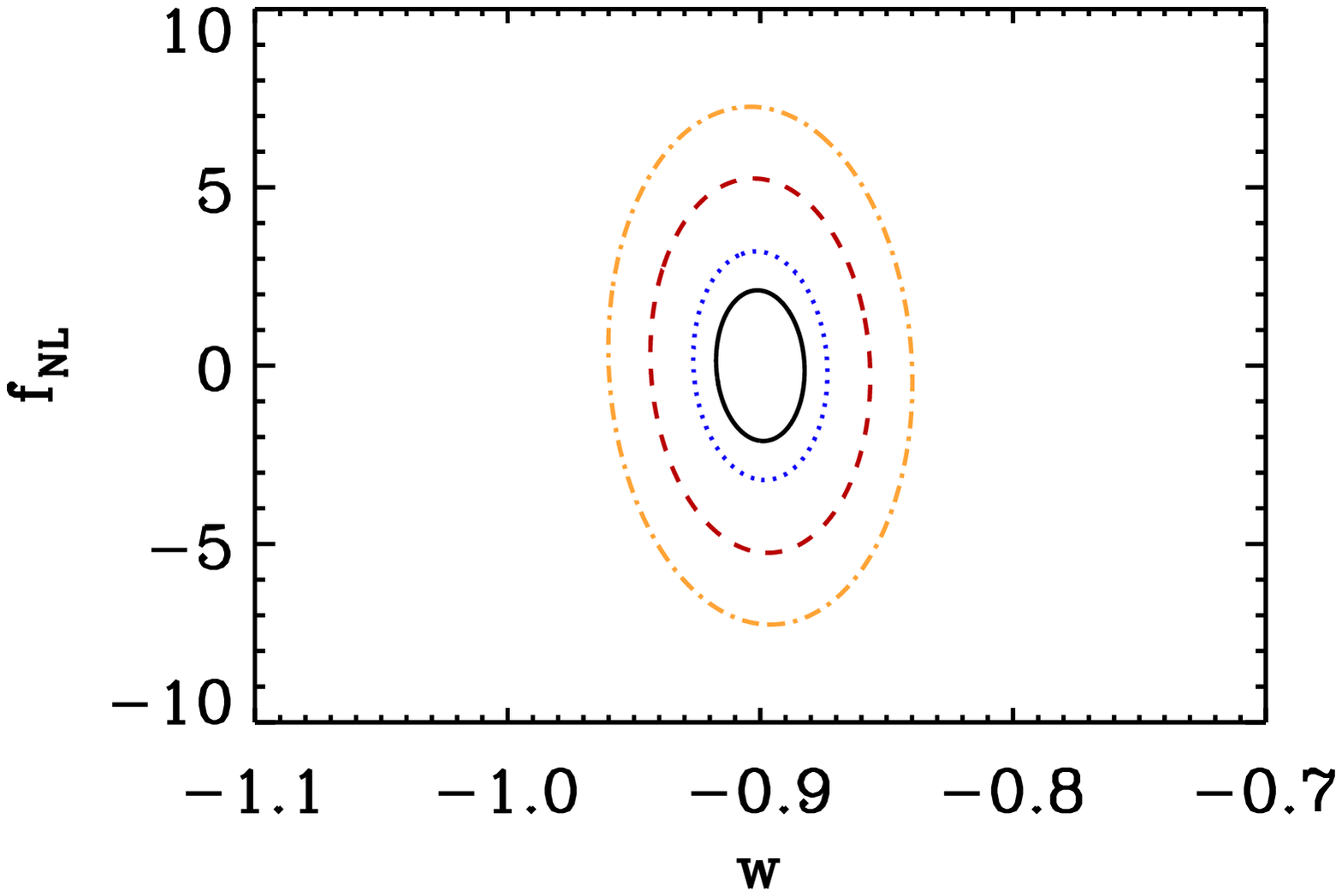}&
\includegraphics[width=3.5cm]{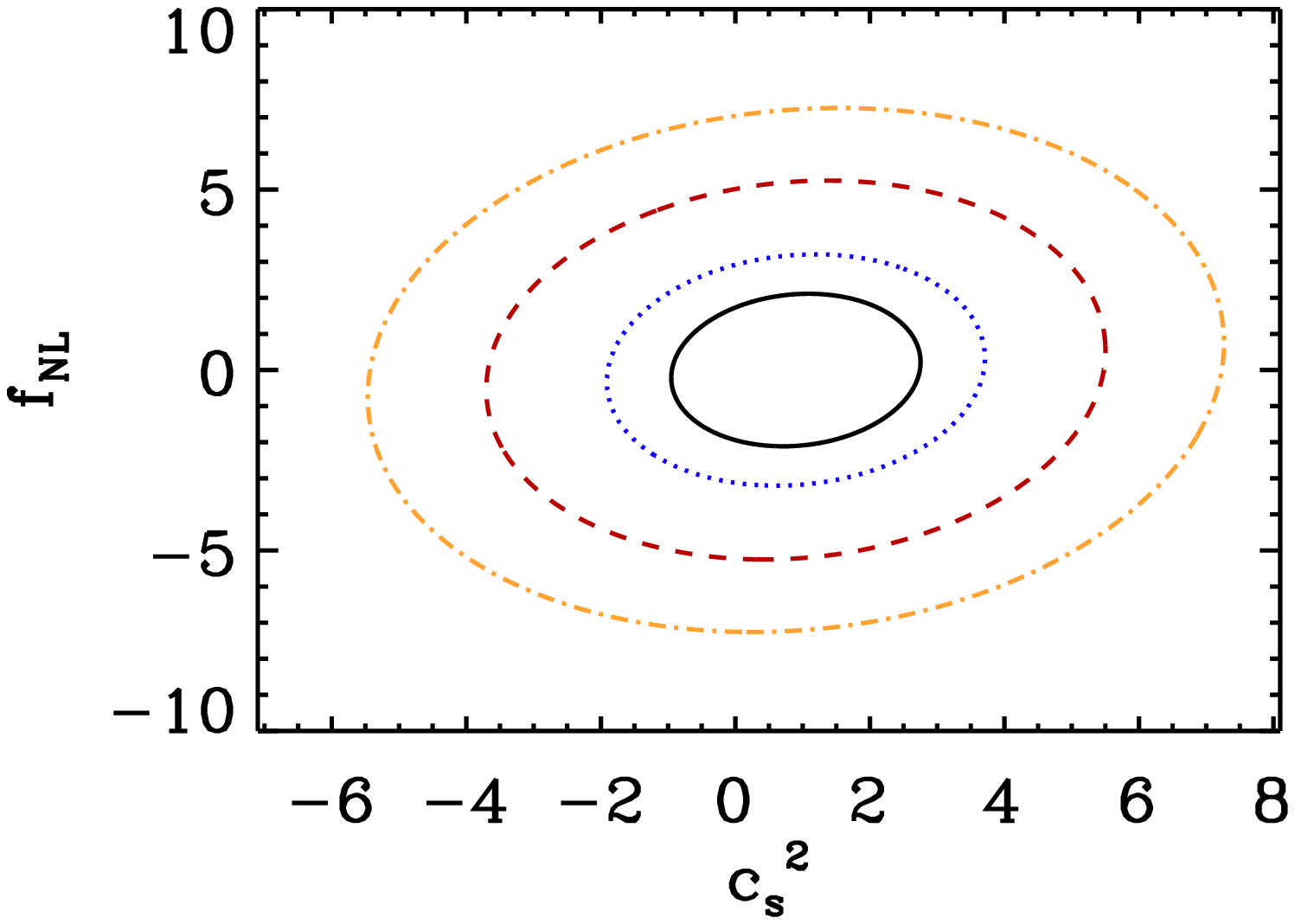}&
\includegraphics[width=3.5cm]{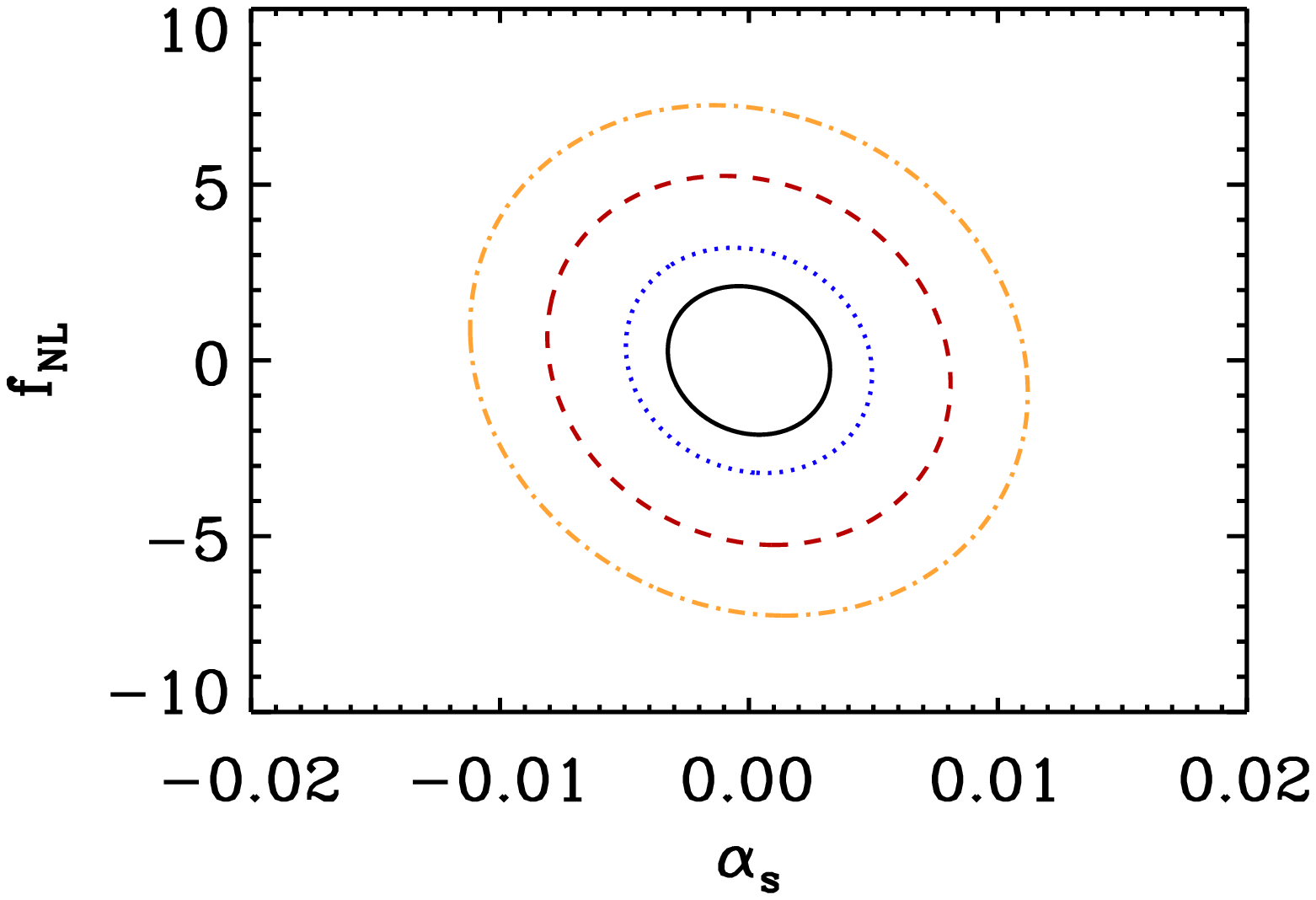}&
\includegraphics[width=3.5cm]{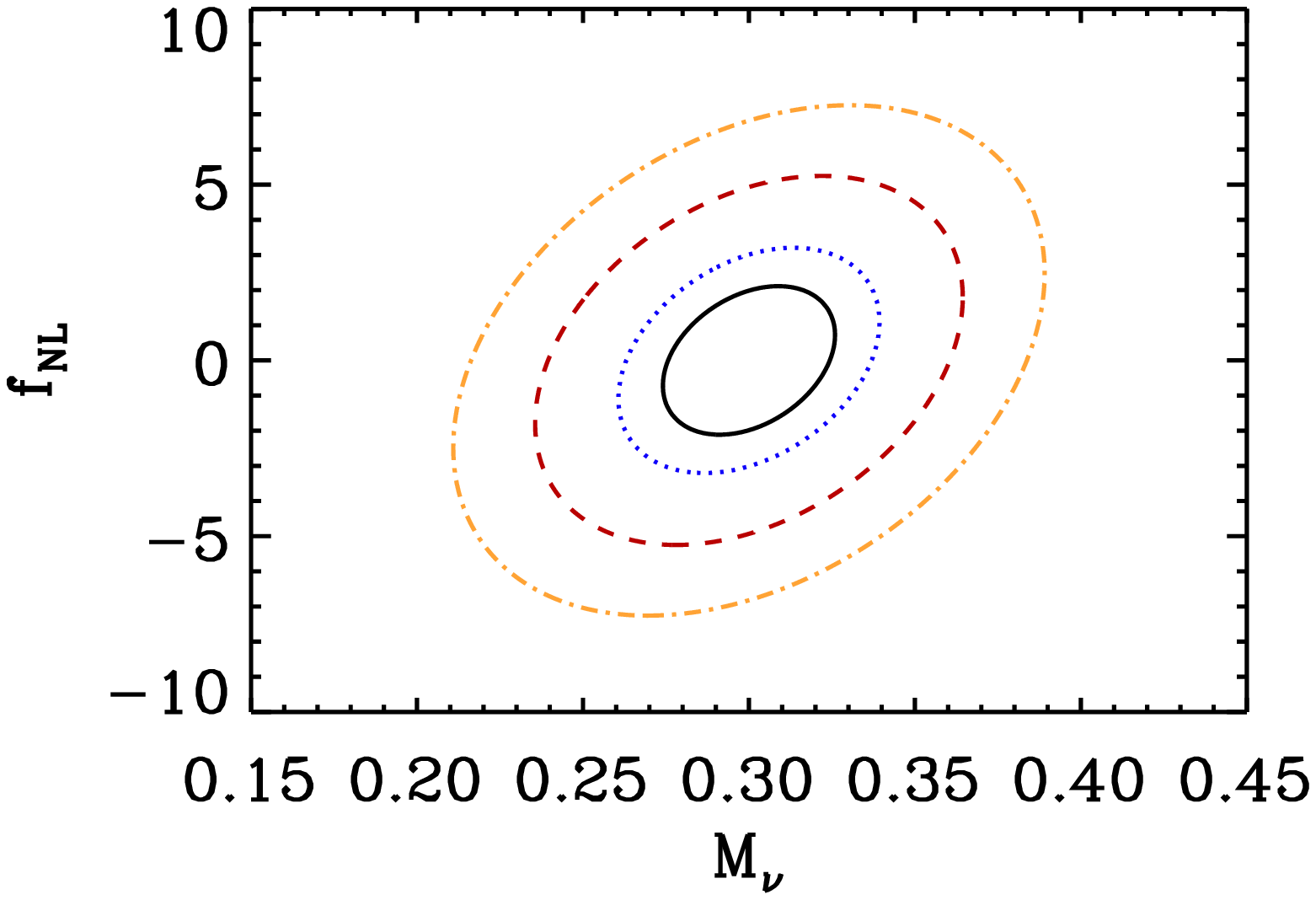}\\
\includegraphics[width=3.5cm]{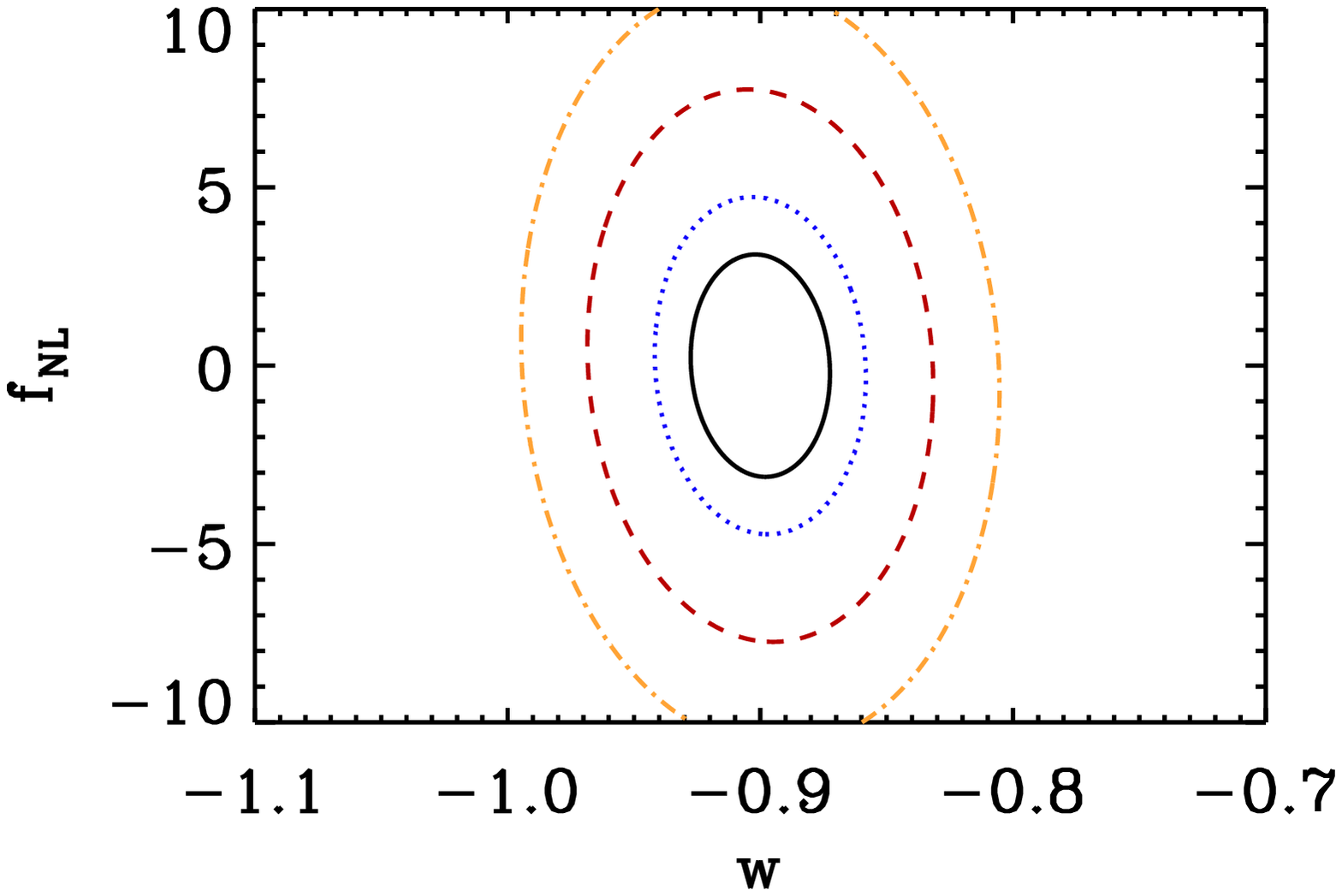}&
\includegraphics[width=3.5cm]{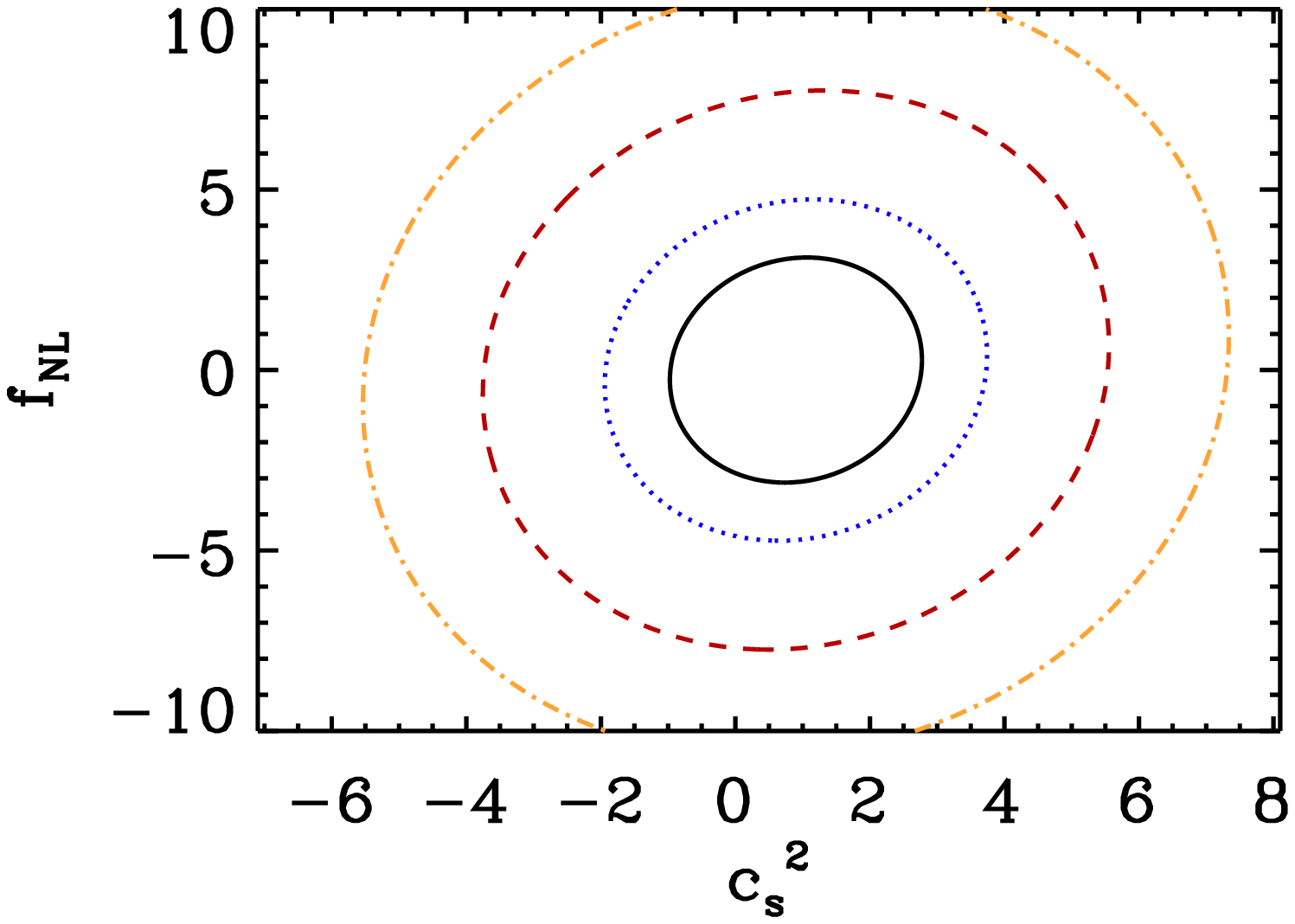}&
\includegraphics[width=3.5cm]{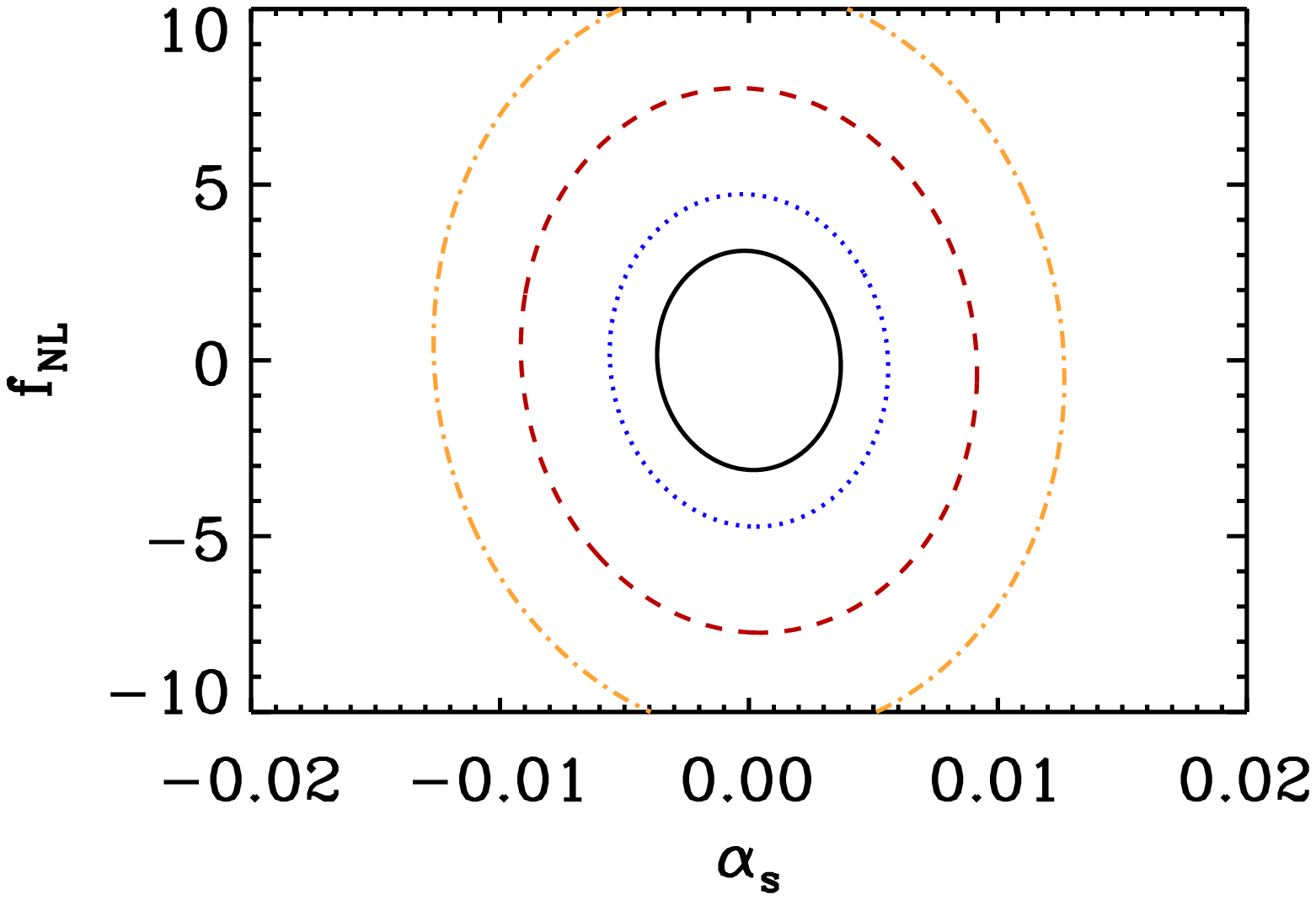}&
\includegraphics[width=3.5cm]{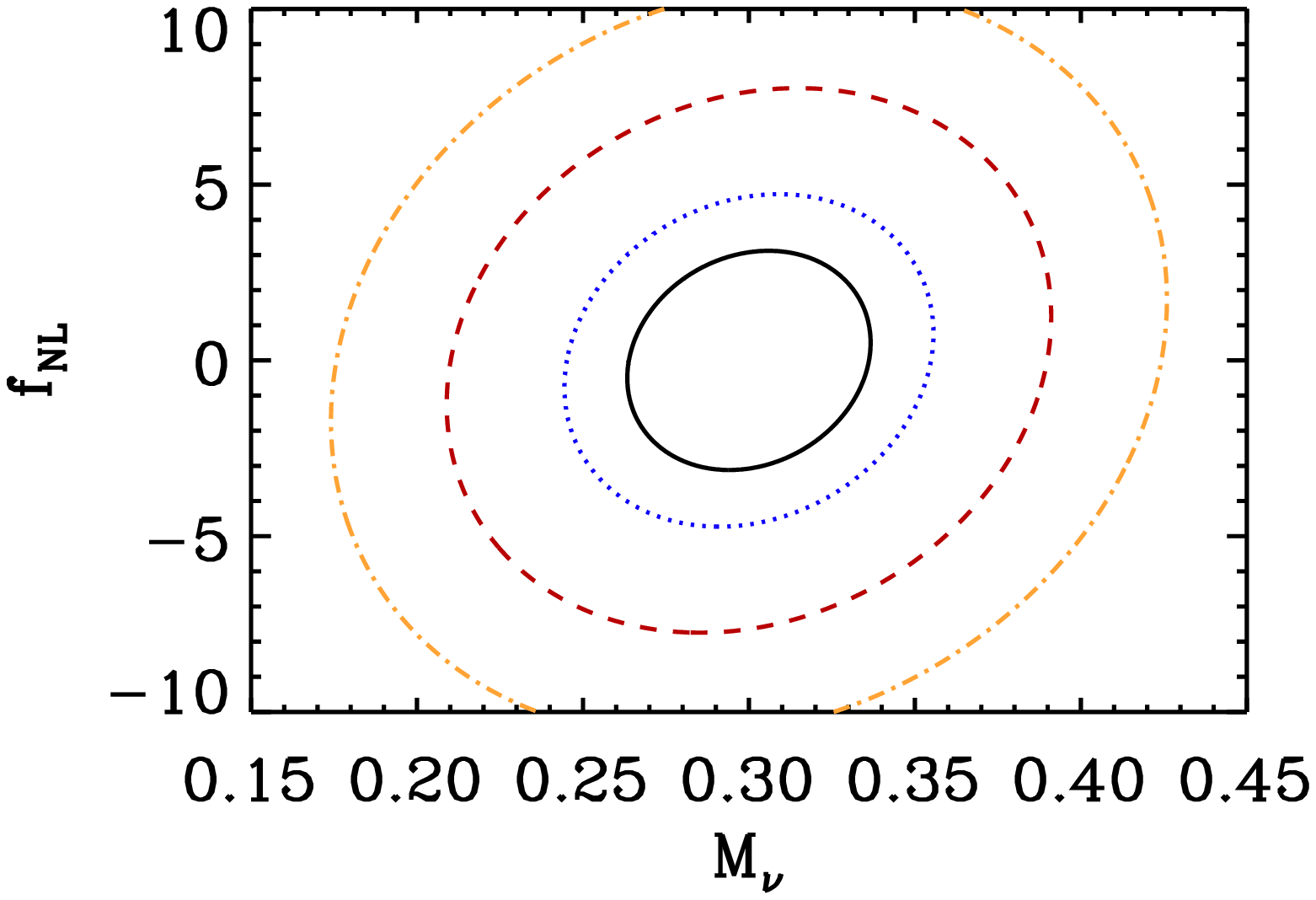}
\end{tabular}
\caption{The same as in Fig.~2, for the fiducial model with
  massive neutrinos of total mass $M_\nu|_{\rm fid}=0.3$ eV, as described in the
  text.}
\label{fig_mnu}
\end{figure}

In Figs.~\ref{fig_Neff}-\ref{fig_mnu} we show the 2-parameter
projected 68$\%$
C.L., 95.4$\%$ C.L. and 99.73$\%$ C.L. contours
in the $f_{\rm NL}$-$p_\alpha$ sub-space with $p_\alpha=w,
c_s^2,\alpha_s, M_\nu, N_\nu^{\rm rel}$, 
obtained
after combining LSST and EUCLID data with Planck priors for the two
fiducial models considered in this work.
The black line
shows the 1-parameter confidence level at 1--$\sigma$.
The orientation of the ellipses reflects the correlations among the
parameters shown in Tabs. \ref{correlation_mnu0.3}-\ref{correlation_Nnu}.

\section{Conclusions}
\label{Conclusions}
Deviations from non-Gaussianity, usually parameterized by the parameter $f_{\rm NL}$, offer a powerful tool to identify the mechanism
which generates the seeds for the structures we observe currently
in our Universe. 

Here we study the impact
of the uncertainties of the cosmological parameters on the $f_{\rm NL}$ errors expected for the case of local non-Gaussianity for the  large-scale non-Gaussian halo bias effect.
We forecast the
correlations among $f_{\rm NL}$ and
the remaining cosmological parameters (including the running of the
spectral index $\alpha_s$, and the dark energy
parameters $w$ and $c^2_s$) within two possible  cosmological models.
The first model contains massive neutrinos (hypothesis robustly confirmed by neutrino
oscillation data), where the total neutrino mass is a parameter to be constrained by the cosmological data. The second  model  assumes massless neutrinos (or neutrinos with a mass too small to be relevant for the cosmological observations considered here) and  allows for extra
relativistic degrees of freedom $N_\nu^{rel}$, which could be induced
by the presence of sterile neutrinos, non minimally coupled
quintessence fields, or even by the violation of the spin statistics
theorem in the neutrino sector.

We follow here a conservative approach, assuming that $f_{\rm NL}$ 
is constrained exclusively from the very large scale halo
power spectrum (i.e. we neglect CMB information on $f_{\rm NL}$), and
restrict ourselves to scales $k \leqslant 0.03 h$/Mpc, without
exploiting information  e.g., from BAOs,
which will further reduce degeneracies and forecasted errors. We present
first the Fisher matrix forecasts for $f_{\rm NL}$ assuming
EUCLID- and LSST-like surveys for the two model cosmologies
considered here. Then, we add the Planck Fisher forecasts for the
remaining cosmological parameters to study the impact on the $f_{\rm NL}$ correlations.

The combined errors on $f_{\rm NL}$ do not change
significantly in the presence of a dark energy equation of state, massive neutrinos, 
running of the spectral index, or clustering of dark energy
perturbations, which are the parameters we have particularly focused
on, since they are expected to affect the matter power spectrum on
large scales, and represent the main deviations from a minimal
$\Lambda$CDM model. However, the errors on
$f_{\rm NL}$ are highly affected in the presence of extra relativistic
degrees of freedom $N_\nu^{rel}$.
We find that if $N_\nu^{\rm rel}$ is assumed to be fixed, the effect
of the uncertainties on the other cosmological parameters increases the
error on $f_{\rm NL}$ only by 10 to 30\% depending on the survey.  
If $N_\nu^{\rm rel}$ is considered as a parameter to be
simultaneously constrained from 
the data, then the uncertainty in the underlying cosmology increases the $f_{\rm NL}$ error by $\sim 80\%$.
We thus conclude that, except for the effect of $N_\nu^{\rm rel}$, 
the halo-bias $f_{\rm NL}$ constraints are remarkably robust to uncertainties in the underlying cosmology.

One important point  to discuss is the effect of the (Gaussian) halo
bias, as its value boosts the effect of $f_{\rm NL}$ 
on the halo power spectrum shape, and in our analysis it has been assumed to be known.
The (Gaussian) halo bias depends strongly on the type of halos
selected by the survey --whether they correspond to 
extremely high and rare  peaks in the initial fluctuation field--, and
on their accretion history. 
Errors on $f_{\rm NL}$ may be improved -at least in principle-  by  up to a factor of two by optimizing the choice of tracers. 

The bias factor itself will need to be estimated from the survey, at
the same time as the 
other cosmological parameters; the signal comes from scales much
smaller than those used here, 
where the NG effect on halo bias is completely negligible. We
estimate that  the error on the (Gaussian) 
halo bias  will be of the same order (in \%)  as the error on the
linear  growth factor $f$ as a function of redshift, 
which is forecasted to be $< \sim  10\%$
\cite{Licia2df,White:2008jy}.  Such a
residual uncertainly  will  
therefore  increase the $f_{\rm NL}$ errors reported here  by  at most 10\%.

Let us recall that the purpose of this work 
is to show the main correlations between $f_{\rm NL}$ and the other
cosmological parameters, and to
understand if these degeneracies can degrade
dramatically the $f_{\rm NL}$ errors. We have shown that, after
the combination with Planck constraints on parameters different from
$f_{\rm NL}$, the degeneracies get mostly
broken, independently on the particular cosmological parameter,
even without adding information from smaller scales corresponding to 
$k>0.03h$/Mpc. 
Therefore we conclude that the $f_{\rm NL}$
constraints are very robust against underlying cosmology assumptions.

Finally, future surveys which
provide a large sample of galaxies or galaxy clusters over a volume
comparable to the Hubble volume (LSST, EUCLID) will measure primordial
non-Gaussianity of the local form with a marginalized 1--$\sigma$
error of the order $\Delta f_{\rm NL} \sim 2-5$, after combination with CMB priors for the
remaining cosmological parameters.
These results are competitive with CMB bispectrum constraints
achievable with an ideal CMB experiment $\Delta f_{\rm NL} \sim$few
\cite{YKW07, LiguoriRiotto08}.

\acknowledgments
CC acknowledges the support from the Agenzia
Spaziale Italiana (ASI, contract N. I/058/08/0).
OM is supported by a Ram\'on y Cajal contract from the MICINN.
LV is supported by FP7-PEOPLE-2007-4-3-IRG n. 202182 and FP7-IDEAS-Phys.LSS 240117. LV and OM are supported by the MICINN grant AYA2008-03531; LV acknowledges
hospitality of Theory Group, Physics Department, CERN, and of
Astronomy department of Universit\a`a di Bologna, where part of the
work was carried out. 

\bibliographystyle{JHEP}

\end{document}